\documentclass[twocolumn,aps,showpacs,amsmath,superscriptaddress]{revtex4-1}

\usepackage{bm}
\usepackage{amsmath}
\usepackage{amssymb}
\usepackage[usenames,dvipsnames]{color}
\usepackage{graphicx}
\usepackage{dcolumn}
\usepackage{simplewick}
\usepackage{hyperref}
\hypersetup{
  colorlinks=true,
  citecolor=blue,
  linkcolor=blue,
  urlcolor=blue}

\begin{document}

\title{Triple excitations in perturbed relativistic coupled-cluster theory
       and Electric dipole polarizability of groupIIB elements}
\author{S. Chattopadhyay}
\affiliation{Department of Physics and Astronomy, Aarhus University,
             DK-8000 Aarhus C, Denmark
             }
\author{B. K. Mani}
\affiliation{Department of Physics, University of South Florida, Tampa,
             Florida 33620, USA}
\author{D. Angom}
\affiliation{Physical Research Laboratory,
             Ahmedabad - 380009, Gujarat,
             India}

\begin{abstract}

\end{abstract}

\begin{abstract}

   We use perturbed relativistic coupled-cluster (PRCC) theory to compute
the electric dipole polarizabilities $\alpha$ of Zn, Cd and Hg. The 
computations are done using the Dirac-Coulomb-Breit Hamiltonian with Uehling 
potential to incorporate vacuum polarization corrections. The triple 
excitations are included perturbatively in the PRCC theory, and in the 
unperturbed sector, it is included non-perturbatively. Our results of $\alpha$, 
for all the three elements, are in excellent agreement with the experimental 
data. The other highlight of the results is the orbital energy corrections
from Breit interactions. In the literature we could only get the data of
Hg \cite{lindroth-89a} and are near perfect match with our results.
We also present the linearized equations of the cluster amplitudes, including 
the triple excitations, with the angular factors. 
\end{abstract}



\maketitle


\section{Introduction}

 Electric dipole polarizability $\alpha$ of atoms, and ions is an important 
property to quantify the response to an external electromagnetic field
\cite{bonin-97}. It is essential to have accurate values of $\alpha$ for atoms, 
and ions in numerous state of the art experiments to probe 
fundamental physics, and develop new technologies. An important example is
the accurate predictions of black-body radiation shift \cite{safronova-12a} 
in optical atomic clocks \cite{diddams-04}, which has been realized with 
optical lattice \cite{takamoto-05}, trapped ions \cite{diddams-01} and 
ultracold atoms \cite{wilpers-02}. In theoretical atomic structure, and 
properties calculations $\alpha$ serves as an excellent proxy to assess the 
accuracy of theoretical many-body calculations. In the present work, the
studies on the $\alpha$ of Hg serves as an appraisal of the many-body
effects important for accurate structure, and properties calculations. This
is a prerequisite to study the permanent electric dipole moment of Hg
\cite{griffith-09} as a signature of parity- and time-reversal violations, and 
probe physics beyond the standard model of particle physics. Given the 
importance of $\alpha$, it has been studied using a variety of many-body 
methods, and are discussed in a recent review by Mitroy and collaborators 
\cite{mitroy-10}. Another reference we have found extremely valuable for our 
studies on the $\alpha$ of neutral atoms is the Schwerdtfeger's updated 
Table of $\alpha$ \cite{note1}, which originally appeared in the chapter
by the same author in the collected volume by 
Maroulis \cite{schwerdtfeger-06}.  The table provides an exhaustive list of 
references on experimental, and theoretical results of $\alpha$ for the 
electronic ground states of neutral elements.

  In the present work we study the $\alpha$ of Zn, Cd and Hg using the
perturbed relativistic coupled-cluster (PRCC) theory. It is built
upon the coupled-cluster theory (CCT), first developed to address the nuclear
many-body problem \cite{coester-58,coester-60}, and later applied to studies
on atom and molecules \cite{cizek-69} . The CCT, and relativistic
version, relativistic coupled-cluster (RCC), are now extensively
used in atomic \cite{pal-07, mani-09,nataraj-11}, molecular \cite{isaev-04}, 
nuclear \cite{hagen-14}, and condensed matter physics \cite{li-14} many-body
calculations. In the PRCC theory, we add a second set of coupled-cluster 
amplitudes to account for an additional interaction Hamiltonian. The method
is general, and can be adapted with ease to incorporate different forms
of interaction Hamiltonians. The detailed descriptions of the theory is 
provided in a series of our previous works \cite{chattopadhyay-12a, 
chattopadhyay-12b, chattopadhyay-13a, chattopadhyay-13b,chattopadhyay-14a}.
Besides the description of PRCC theory, through these works we had 
explored the impact of Breit interaction \cite{chattopadhyay-12b}, 
improved diagrammatic evaluations \cite{chattopadhyay-13a},
vacuum polarization \cite{chattopadhyay-13b}, and triple excitation cluster 
operators \cite{chattopadhyay-14a} in the unperturbed cluster operators.
A related method used for calculating electric dipole polarizabilities is to 
consider the $z$-component of the dipole operator and define a set of 
perturbed cluster operators \cite{sahoo-08,yashpal-13}. In the present work, 
we report the inclusion of the dominant perturbative triples in 
the PRCC theory, and improved validation of including Breit interaction in 
the generation of orbital basis set and PRCC theory. Our earlier works, related
to Breit interaction, reported matching the Dirac-Coulomb-Breit SCF energies 
with previous results. This, however, provides an assessment of the 
implementation at a coarse grained level. A better comparison would be the 
orbital energy corrections from the Breit interaction. This is what we 
demonstrate for Hg, as we could get the data  from a previous 
work \cite{lindroth-89a}. This, we feel, is an important validation of our 
implementation of Breit interaction. 

  The important feature of the present work is, it extends, and verify the 
applicability of PRCC theory in the computation of $\alpha$ to the transition
elements. As expected, we get very good results, and  we have gained 
significant insight on the nature of the correlation effects with $d$ 
sub-shell as the immediate shell below the valence.
 
  The remaining part of the paper consists of five sections. In next section,
Section II, we provide a brief discussion on the RCC theory. The description
of the linearized RCC and PRCC equations, along with the angular factors,
form the principal parts of subsections in this section. It must be 
emphasized that the linearized RCC equations include the triple excitation
cluster amplitudes, with the representation we introduced in our previous 
work \cite{chattopadhyay-14a}. The Section III provides a brief description
of how to compute $\alpha$ with PRCC, and is followed with an exposition on
the computational details in Section IV. The results and discussions are
given in Section V. We provide detailed analysis of our theoretical results,
and discuss, vis-a-vis previous results, relevant trends and prospects for
possible future improvements. We, then, end the main part of the paper with
conclusions. In the appendix, we have listed the angular factors of the
linearized RCCSDT and PRCC. With these, we feel, interested readers would
be able to implement these theories at the linear level without difficulty.
For the details on the nonlinear terms, the readers may refer our previous 
work \cite{chattopadhyay-12b}.  The results and equations presented in 
this work are in atomic units ( $\hbar=m_e=e=1/4\pi\epsilon_0=1$). In 
this system of units the velocity of light is $\alpha ^{-1}$, the inverse 
of fine structure constant. For which we use the value 
of $\alpha ^{-1} = 137.035\;999\;074$ \cite{mohr-12}.


\section{Relativistic Coupled-cluster theory}

  The Dirac-Coulomb-Breit Hamiltonian $H^{\rm DCB}$ provides a good description
of neutral atom, and well suited for structure and properties calculations.
For an $N$-electron atom
\begin{eqnarray}
   H^{\rm DCB} & = & \Lambda_{++}\sum_{i=1}^N \left [c\bm{\alpha}_i \cdot 
        \mathbf{p}_i + (\beta_i -1)c^2 - V_{N}(r_i) \right ] 
                       \nonumber \\
    & & + \sum_{i<j}\left [ \frac{1}{r_{ij}}  + g^{\rm B}(r_{ij}) \right ]
        \Lambda_{++},
  \label{ham_dcb}
\end{eqnarray}
where $\bm{\alpha}$ and $\beta$ are the Dirac matrices, $\Lambda_{++}$ is an 
operator which projects to the positive energy solutions and $V_{N}(r_{i})$ is 
the nuclear potential. Sandwiching the Hamiltonian with $\Lambda_{++}$ ensures 
that the effects of the negative energy continuum  states are neglected in the 
calculations. Another approach, which is better suited for numerical
computations, is to use the kinetically balanced finite basis sets
\cite{stanton-84,mohanty-90,grant-06,grant-10}. We  use this method in the
present work to generate the orbital basis sets.
The last two terms, $1/r_{ij} $ and $g^{\rm B}(r_{ij})$  are the 
Coulomb and Breit interactions, respectively.  The later, Breit interaction, 
represents the inter-electron magnetic interactions and is given by
\begin{equation}
  g^{\rm B}(r_{12})= -\frac{1}{2r_{12}} \left [ \bm{\alpha}_1\cdot\bm{\alpha}_2
               + \frac{(\bm{\alpha_1}\cdot \mathbf{r}_{12})
               (\bm{\alpha_2}\cdot\mathbf{r}_{12})}{r_{12}^2}\right].
\end{equation}
The Hamiltonian satisfies the eigen-value equation
\begin{equation}
   H^{\rm DCB}|\Psi_{i}\rangle = E_{i}|\Psi_{i}\rangle , 
\end{equation}
where, $|\Psi_{i}\rangle$ is the exact atomic state and $E_i$ is the energy 
of the atomic state. In the presence of external electromagnetic fields, the
Hamiltonian is modified with the addition of interaction terms. For external 
static electric field, the interaction is 
$H_{\rm int}=-\mathbf{d}\cdot\mathbf{E}_{\rm ext} $, 
where $\mathbf{d}$ and $\mathbf{E}_{\rm ext}$ are the induced electric dipole 
moment of the atom and external electric field, respectively. In the remaining
part of this section we give a brief description of RCC theory, which we use
to compute atomic state $|\Psi\rangle $ and PRCC to account for the 
effects of $H_{\rm int}$ in the atomic state. 


\subsection{Overview of RCC and PRCC theories}

 In RCC theory we define the  ground state atomic wavefunction of a 
closed-shell atom as 
\begin{equation}
|\Psi_0\rangle = e^{ T^{(0)}}|\Phi_0\rangle ,
\end{equation}
where $|\Phi_0\rangle$ is the reference state wave-function and $T^{(0)}$ is 
the unperturbed cluster operator. The wave-function is modified when the atom
is subjected to an external static electric field $\mathbf{E}$, and the 
interaction Hamiltonian is $ H_{\rm int} = -\mathbf{D}\cdot\mathbf{E}$,
where $\mathbf{D}$ is the induced electric dipole moment of the atom. 
In the present work we define the perturbed ground state as
\begin{equation}
 |\tilde{\Psi}_0\rangle = e^{T^{(0)} + \lambda \mathbf{T}^{(1)}\cdot\mathbf{E}} 
 |\Phi_0\rangle = e^{T^{(0)}}\left [ 1 + \lambda \mathbf{T^{(1)}\cdot 
 \mathbf{E}} \right ] |\Phi_0\rangle , \;\;\;\;
 \label{psi_tilde}
\end{equation}
where $\mathbf{T}^{(1)}$ are the PRCC operators 
\cite{chattopadhyay-12a,chattopadhyay-12b}. For an $N$-electron 
closed-shell atom $T^{(0)} = \sum_{i=1}^N T_{i}^{(0)}$ and 
${\mathbf T}^{(1)} = \sum_{i=1}^N {\mathbf T}_{i}^{(1)}$, where $i$ is the 
order of excitation.  In the coupled-cluster single and double (CCSD) 
excitation approximation \cite{purvis-82},
\begin{subequations}
\begin{eqnarray}
T^{(0)} &=& T_{1}^{(0)} + T_{2}^{(0)}, \\ 
{\mathbf T}^{(1)} &=& {\mathbf T}_{1}^{(1)} + {\mathbf T}_{2}^{(1)}.
\end{eqnarray}
\end{subequations}
The CCSD is a good starting point for structure and properties calculations 
of closed-shell atoms and ions. In the second quantized representation
\begin{subequations}
\begin{eqnarray}
  T_1^{(0)} &= &\sum_{a,p} t_a^p {{a}_p^\dagger} a_a , 
                \label{t1_def}         \\
  T_2^{(0)} &= &\frac{1}{4}\sum_{a,b,p,q}
                t_{ab}^{pq} {{a}_p^\dagger}{{a}_q^\dagger}a_b a_a , 
                \label{t2_def}         \\
  \mathbf{T}_1^{(1)} & = & \sum_{a,p} \tau_a^p \mathbf{C}_1 (\hat{r})
                       a_{p}^{\dagger}a_{a},
                \label{pt1_def}         \\
  \mathbf{T}_2^{(1)} & = & \frac{1}{4}\sum_{a,b,p,q} \sum_{l,k} 
                    \tau_{ab}^{pq}(l,k) \{ \mathbf{C}_l(\hat{r}_1) 
                    \mathbf{C}_k(\hat{r}_2)\}^{1}
                   a_{p}^{\dagger}a_{q}^{\dagger}a_{b}a_{a}, \;\;\;\;\;\;\;\;
                \label{pt2_def}  
\end{eqnarray}
\end{subequations}
where $t_{\ldots}^{\ldots}$ and $\tau_{\ldots}^{\ldots}$ are the cluster 
amplitudes, $a_i^{\dagger}$ ($a_i$) are single particle creation (annihilation)
operators and $abc\ldots$ ($pqr\ldots$) represent core (virtual) single 
particle states or orbitals. To represent $\mathbf{T}_1^{(1)}$, a rank one 
operator, we have used the $\mathbf{C}$-tensor of similar rank 
$\mathbf{C}_1(\hat r)$. Coming to $\mathbf{T}_2^{(1)}$, to represent it two 
$\mathbf{C}$-tensor operators of rank $l$ and $k$ are coupled to a rank one 
tensor operator. In addition, the PRCC clusters are constrained by 
other selection rules arising from parity and triangular conditions, 
these are described in our previous work \cite{chattopadhyay-12b}.

With the inclusion of $T_3^{(0)}$ the RCC theory incorporates all the 
correlation effects  up to second order in the residual Coulomb interaction. 
That is, the theory encapsulates all the many-body perturbation theory (MBPT) 
diagrams \cite{lindgren-86} which are first and second order in the 
residual Coulomb interaction. In addition, as it is coupled cluster theory, 
it incorporates the connected single, double and triple excitations 
to all order. The leading order contribution to the uncertainty in the 
calculations arise from the quadruple excitations, which, in MBPT, first 
appear at the third order of perturbation.


\subsection{Linearized CCSDT cluster equations}

 A simplified approximation which incorporates most of the important the 
many-body effects is the linearized RCCSDT. In this approximation we
only consider terms which are zeroth and first order in the cluster 
operators. The importance of the linearized cluster equations is that, to 
solve the RCCSDT equations iteratively, we take the solutions 
as the initial values. The $T_1^{(0)}$, $T_2^{(0)}$ and 
$T_3^{(0)}$ cluster equations, as described in our previous work, are then
\begin{widetext}
\begin{eqnarray}
   \label{lccsdt_s}
   & & \sum_{bq} g^{bp}_{aq}t_b^q + \frac{1}{2}\sum_{bcq}g^{bc}_{qa}
       (t_{bc}^{qp} - t_{bc}^{pq}) 
       + \sum_{bqr}g^{bp}_{qr}(t_{ba}^{qr} - t_{ab}^{qr}) 
       + \frac{1}{2}\sum_{bcqr} (g^{bc}_{qr} - g^{bc}_{rq})t_{abc}^{pqr} 
       + \left (\varepsilon_p - \varepsilon_a \right ) t_a^p= 0, \\
   & & \sum_{r}g^{pq}_{ar}t_b^r - \sum_{c}g^{pc}_{ab}t_c^q 
       + \sum_{cd}g^{cd}_{ab}t_{cd}^{pq} + \sum_{rs}g^{pq}_{rs}t_{ab}^{rs} 
       - \sum_{cr} \bigg [ g^{cp}_{ar} t_{cb}^{rq}   
       + g^{pc}_{rb}t_{ac}^{rq} + \frac{1}{2} g^{pc}_{ar}(t_{cb}^{rq} 
       - t_{bc}^{rq}) \bigg ] + \sum_{rcs}(g^{rs}_{cq} 
       - g^{sr}_{cq}) t_{acb}^{prs} 
       \nonumber \\
   \label{lccsdt_d}
   & & + \frac{1}{2} \sum_{rcd}(g^{rb}_{cd}
       - g^{rb}_{dc})t_{acd}^{prq}  +  \left ( \begin{array}{c}
                                                   p \leftrightarrow q \\
                                                   a \leftrightarrow c
                                                 \end{array}\right )
       + \left (\varepsilon_p + \varepsilon_q 
       - \varepsilon_a - \varepsilon_b\right )t_{ab}^{pq} + g_{ab}^{pq}= 0, \\
   & & \sum_{s} g^{qr}_{sc}t_{ab}^{ps}   + \sum_{d} g^{dr}_{bc}t_{ad}^{pq}
       + \sum_{ds} \bigg [ g^{as}_{pd}\left ( t_{dbc}^{sqr} 
       + t_{bdc}^{sqr} \right )  + g^{sb}_{pd}t_{adc}^{sdr}
       + g^{as}_{dp}t_{dbc}^{sqr} \bigg ] 
       + \sum_{st} g^{st}_{pq}t_{abc}^{str} + \sum_{de} g^{ab}_{de}t_{dec}^{pqr}
       + \left ( \begin{array}{c}
                    p \leftrightarrow q \leftrightarrow r\\
                    a \leftrightarrow b \leftrightarrow c
                 \end{array}\right )
       \nonumber \\
   & & + \left (\varepsilon_p + \varepsilon_q + \varepsilon_r- \varepsilon_a 
    - \varepsilon_b - \varepsilon_c\right ) t_{abc}^{pqr}= 0.
   \label{lccsdt_t}
\end{eqnarray}
\end{widetext}
where, $\varepsilon_i$ is the orbital energy of the $i$th orbital,
$i\leftrightarrow j$ represents permutation of the two indexes and 
$g_{ij}^{kl} = \langle kl|1/r_{12} + g^{B}(r_{12})|ij\rangle$ is the matrix
element of the two-electron interaction Hamiltonian. For the 
cluster amplitudes $t_{abc}^{pqr}$, we use the representation introduced in 
our previous work \cite{chattopadhyay-14a}. The representation is symmetric 
with respect to the interchange of orbital indexes and reduces the number of 
terms in the equations. So, in the cluster equations, only classes of 
contractions based on the number of hole (particle) are considered or terms 
with unique topology of the Goldstone diagrams are considered in the 
equations. Another equivalent representation of $t_{abc}^{pqr}$ with a 
different multipole structure is given in the work of Derevianko and 
collaborators \cite{derevianko-08}.

The Eqs. (\ref{lccsdt_s}-\ref{lccsdt_t}) are in terms of the matrix elements 
of the two-electron interactions. Another representation which is suitable
for atomic or ionic systems, and consistent with the expressions in 
properties calculations is to write the equations in terms of reduced matrix
elements. For this consider the matrix element of the 
electron-electron Coulomb interaction, following the standard multipole 
decomposition \cite{lindgren-86,johnson-07,grant-10}
\begin{eqnarray}
  \langle pq|\frac{1}{r_{12}}|ab\rangle & = &\sum_k \sum_q 
    \left ( \begin{array}{ccc}
              j_p & k & j_a \\
             -m_p & q & m_a \\
            \end{array} \right )
    \left ( \begin{array}{ccc}
              j_q &  k & j_b \\
             -m_q & -q & m_b \\
            \end{array} \right ) \nonumber \\
   && \times X^k_{\rm C}(pqab),
\end{eqnarray} 
where, we have followed the notations in Ref. \cite{grant-10}. In the 
above expression $X^k_{\rm C}(pqab)$ is the reduced matrix element or the part 
of the matrix element which is independent of the magnetic quantum numbers. It 
is defined as
\begin{eqnarray}
  X^k_{\rm C}(pqab) &= & \{j_p,j_a,k\}\{j_q,j_b,k\}
      \Pi^e(\kappa_p \kappa_a k) \Pi^e(\kappa_q \kappa_b k)  \;\;\;\;\;
      \nonumber \\
  & & (-1)^k \langle j_p||\mathbf{C}^k||j_a\rangle
      \langle j_q||\mathbf{C}^k||j_b\rangle R^k_{\rm C}(pqab),
\end{eqnarray}
where, $\{j_i,j_j,k\} $ is the triangular condition, 
$\Pi^e(\kappa_i \kappa_j k)$ is the parity condition that $l_i + l_j + k$ must
be even, $\mathbf{C}^k$ is a c-tensor and $ R^k_{\rm C}(pqab)$ is the radial 
part of the matrix element. The matrix elements of the Breit iteration, 
$g^{\rm B}(r_{12}) $, may also be written in a similar form. For this
let $X^k_{\rm B}(pqab)$ represents the reduced matrix element of 
$g^{\rm B}(r_{12}) $ and as a compact notation define
\begin{equation}
   g_{ab,k}^{pq} = X^k_{\rm C}(pqab) + X^k_{\rm B}(pqab),
\end{equation}
as the two-electron reduced matrix element corresponding to the multipole $k$.
Based on this definition, the cluster amplitudes must also be defined in 
terms of the multipole structure, and we use the notation $t_{ab,k}^{pq}$ to 
represent the component of $t_{ab}^{pq}$ with multipole $k$. The cluster 
amplitude equations Eqs. (\ref{lccsdt_s}-\ref{lccsdt_t}) are then in terms of 
reduced matrix elements. To examine the multipole structure of $T_2^{(0)}$, 
consider the approximation based on the first order in many-body perturbation 
theory (MBPT). The cluster amplitude is then
\begin{equation}
   t_{ab,k}^{pq} \approx \frac{g_{ab,k}^{pq}}{(\varepsilon_p + \varepsilon_q 
                 - \varepsilon_a - \varepsilon_b)}.
\end{equation} 
It must be mentioned here that this is also the expression we use as the 
initial guess to solve the cluster equations iteratively using a method like 
Jacobi. In this case, the multipoles $k$ of the cluster amplitudes are 
identical to the two-electron interactions. The triple cluster amplitudes 
$t_{abc}^{pqr} $, however, involves three multipoles and details related to 
the multipole representation are discussed in our previous 
work \cite{chattopadhyay-14a}. A similar description on the cluster equations 
with the CCSD approximation, in terms of reduced matrix elements, is presented 
in Ref. \cite{blundell-89}. The reference also provides detailed expressions 
of the angular factors corresponding to each term in the cluster equation. 
Adopting the notations defined here and representations of $t_{abc}^{pqr} $ 
discussed in our previous work \cite{chattopadhyay-14a}, we use 
$t_{abc,l_1l_2l_3}^{pqr}$ to represent the cluster amplitudes
of $T_3^{(0)}$ in terms of reduced matrix elements. Where, $l_i$s are the 
multipoles in the representation of the $T_3^{(0)}$ cluster 
amplitudes \cite{chattopadhyay-14a}. The cluster equations are then
\begin{widetext}
\begin{eqnarray}
   & & \sum_{bq} \left ( A_1 g_{aq,0}^{pb}
       - A_2 g_{aq,0}^{bp} 
         \right ) t_b^q  
       + \sum_{bcqk_1}g^{bc}_{qa,k_1} \left (
         A_3t_{bc,k_1}^{qp} 
         - \sum_{k_2}
                  A_4t_{bc,k_2}^{pq} \right ) 
       + \sum_{bqrk_1}g_{qr,k_1}^{bp}\left (
          A_5t_{ba,k_1}^{qr} 
         - \sum_{k_2}
           A_6t_{ab,k_2}^{qr} \right )
                                    \nonumber   \\
    & &+ \frac{1}{2}\sum_{bcqrl_1} \left ( 
         A_7g_{bc,l_1}^{qr} 
         - 
         A_8g^{rq}_{bc,l_1} \right ) t_{abc,0l_1l_1}^{pqr} 
       + \left (\tilde{\varepsilon}_p 
       - \tilde{\varepsilon}_a \right ) t_a^p= 0,  
   \label{lccsdt_s}  \\
   & & \sum_{r}
         B_1g_{ar,k}^{pq}t_b^r 
       - \sum_{c}
          B_2g_{ab,k}^{pc}t_c^q 
       + \sum_{cdk_1k_2}
         B_3g_{ab,k_1}^{cd}t_{cd,k_2}^{pq} 
       + \sum_{rsk_1k_2}
         B_4g_{rs,k_1}^{pq}t_{ab,k_2}^{rs}  
       - \sum_{cr} \left (
         B_5g_{ar,k_1}^{cp} t_{cb,k}^{rq}
       + \sum_{k_1k_2}
         B_6 g_{rb,k_1}^{pc}t_{ac,k_2}^{rq} \right .
              \nonumber           \\ 
    & &+ \left . 
         B_7g_{ar,k}^{pc}t_{cb,k}^{rq} 
         -\sum_{k_1} 
         B_8g_{ar,k}^{pc}t_{bc,k_1}^{rq} \right )
        + \sum_{rcsl_1l_2}\left (
         B_9g_{rs,l_1}^{cq} 
            - \sum_{k_1}
         B_{10}g_{sr,k_1}^{cq}\right )
            t_{acb,kl_1l_2}^{prs} 
       + \frac{1}{2} \sum_{rcdl_1l_2}\left (
         B_{11} g_{rb,l_1}^{cd} \right .
         \nonumber       \\ 
   & & - \left . \sum_{k_1}
         B_{12}g_{rb,k_1}^{dc} \right )
         t_{acd,kl_1l_2}^{prq}  
       +  \left ( \begin{array}{c}
                                                   p \leftrightarrow q \\
                                                   a \leftrightarrow c
                                                 \end{array}\right )
       + \left (\tilde{\varepsilon}_p
                + \tilde{\varepsilon}_q
                - \tilde{\varepsilon}_a
                - \tilde{\varepsilon}_b\right )t_{ab,k}^{pq} 
       + g_{ab,k}^{pq}= 0,
   \label{lccsdt_d} \\ 
   & & \sum_{s} C_1
               g_{sc,l_3}^{qr}t_{ab,l_1}^{ps}   
       + \sum_{d} C_2
                g_{bc,l_3}^{dr}t_{ad,l_1}^{pq}
       + \sum_{ds} \bigg [ g_{as,l_1}^{pd}\left ( C_3
                t_{dbc,l_1l_2l_3}^{sqr} \right .
       + \left . C_4
            t_{bdc,m_1m_2l_3}^{sqr} \right )  
       + \sum_{m_1m_2k}C_5
               g_{sb,k}^{pd}t_{adc,m_1m_2l_3}^{sdr}
                             \nonumber \\
   & & + \sum_{k}C_6
               g_{as,k}^{dp}t_{dbc,l_1l_2l_3}^{sqr} \bigg ] 
       + \sum_{st}\sum_{m_1m_2k} C_7
               g_{st,k}^{pq}t_{abc,m_1m_2l_3}^{str} 
       + \sum_{de}\sum_{m_1m_2k}  C_8
         g_{ab,k}^{de}t_{dec,m_1m_2l_3}^{pqr}
       + \left ( \begin{array}{c}
                    p \leftrightarrow q \leftrightarrow r\\
                    a \leftrightarrow b \leftrightarrow c
                 \end{array}\right )
       \nonumber \\
   & & + \left (\tilde{\varepsilon}_p 
              + \tilde{\varepsilon}_q 
              + \tilde{\varepsilon}_r
              - \tilde{\varepsilon}_a 
              - \tilde{\varepsilon}_b 
              - \tilde{\varepsilon}_c\right ) t_{abc}^{pqr}= 0,
   \label{lccsdt_t}
\end{eqnarray}
\end{widetext}
where $A_i$, $B_i$ and $C_i$ are the angular factors given in Appendix
\ref{app_a}-\ref{app_c}, and 
$\tilde{\varepsilon}_i=\epsilon/\sqrt{[j_i]}$ with $[j_i] = 2j_i + 1$. These
are the cluster amplitude equations we solve in the LCCSDT theory.


\subsection{Linearized PRCC equations}

\begin{figure}[h]
 \includegraphics[width=8.5cm]{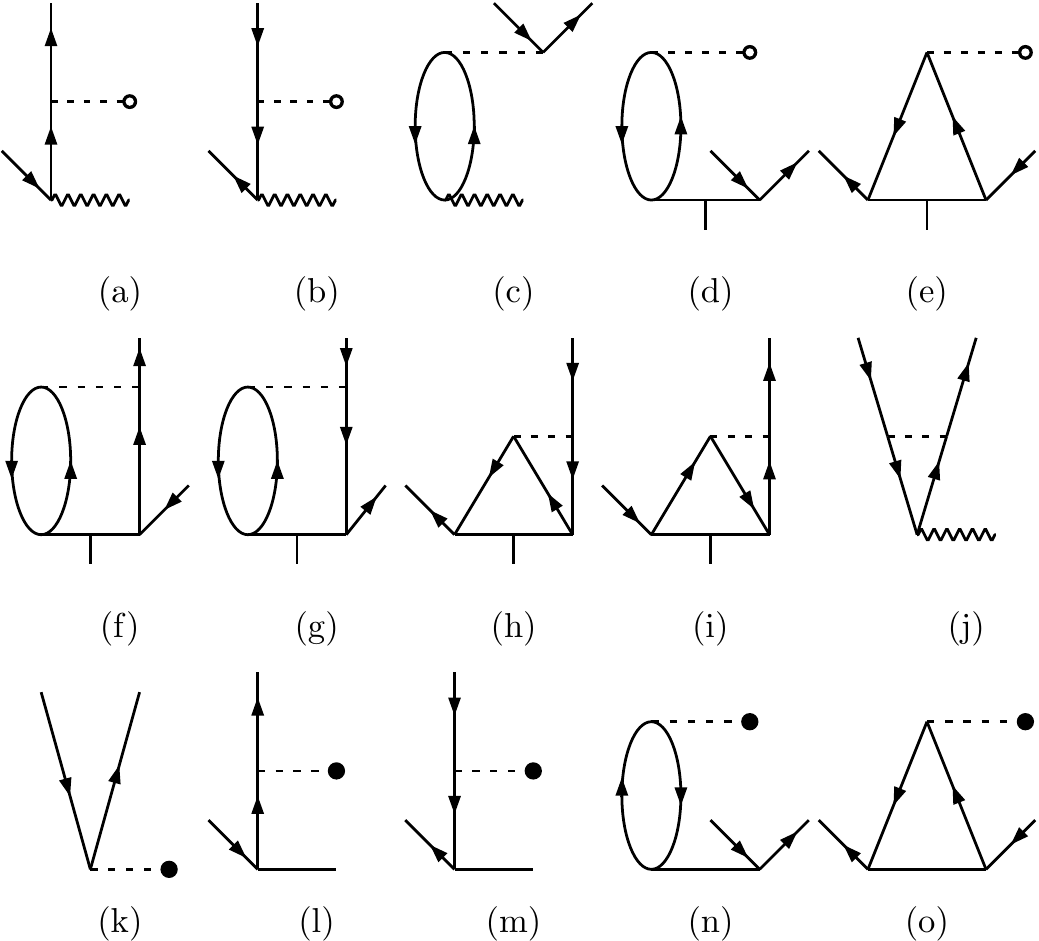}
 \caption{Goldstone diagrams which contribute to the $\mathbf{T}_1^{(1)}$
          equation in the LPRCC approximation. The diagrams (a-c), and (j)
          arise from ${H}_{\rm N}\mathbf{T}^{(1)}_1$, 
          (d-i) arise from ${H}_{\rm N}\mathbf{T}^{(1)}_2$, and (k-o) arise 
          from $\mathbf{d}\mathbf{T}^{(1)}$. The dashed lines ending with a 
          circle ($\circ$) and filled circle ($\bullet$) correspond to 
          interactions associated with the single-body part of $H_N$ and 
          $H_{\rm int}$, respectively. The vertexes with undulating line
          and a short vertical stump represent $\mathbf{T}_1^{(1)}$ and
          $\mathbf{T}_2^{(1)}$, respectively.
         }
 \label{psing_diag}
\end{figure}

\begin{figure}[h]
 \includegraphics[width=8.5cm]{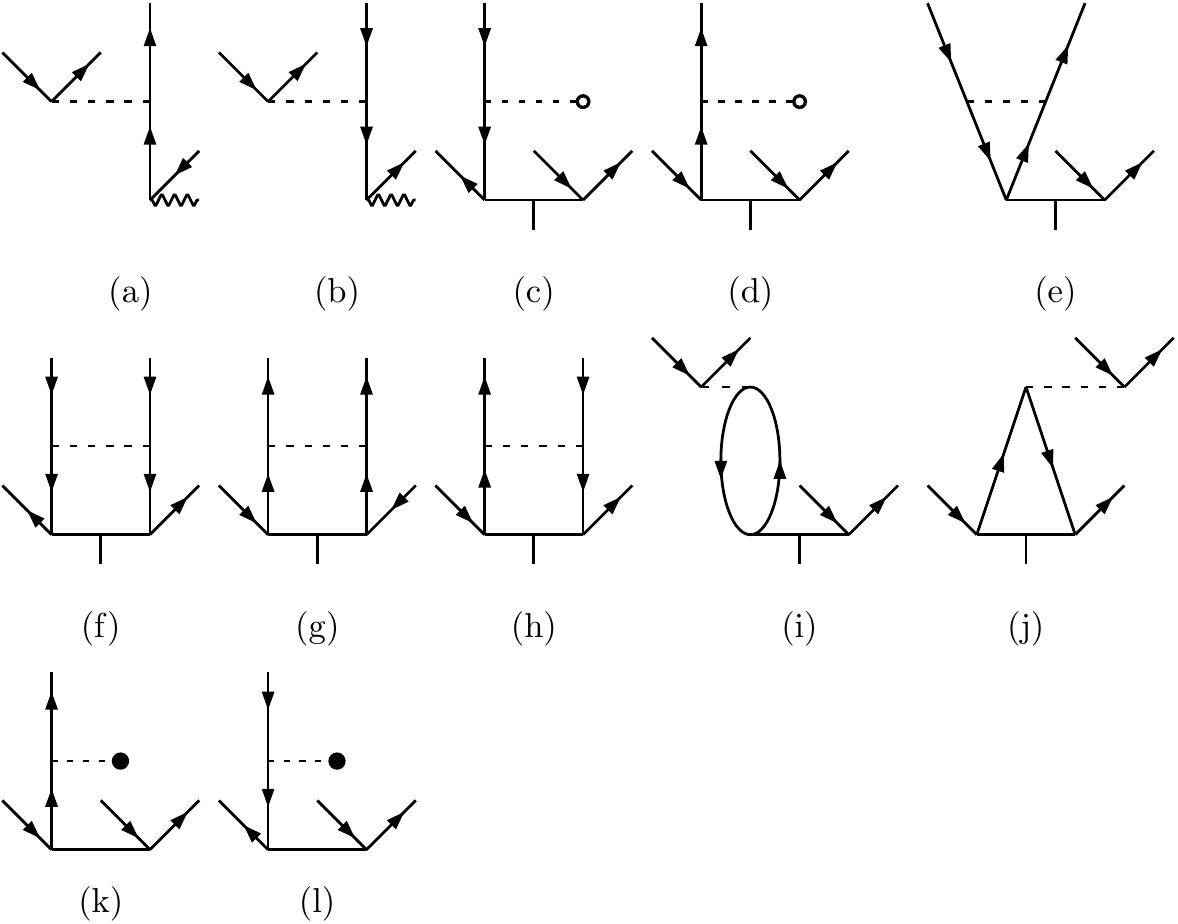}
\caption{Goldstone diagrams which contribute to the $\mathbf{T}_2^{(1)}$
         in the LPRCC approximation. The diagrams (a-b), (c-j), and
         (k-l) arise from ${H}_{\rm N}\mathbf{T}^{(1)}_1$, 
         ${H}_{\rm N}\mathbf{T}^{(1)}_2$, and ${\mathbf{d}}\mathbf{T}^{(1)}_2$, 
         respectively. The dashed lines ending with a circle ($\circ$) and 
         filled circle ($\bullet$) correspond to interactions associated with 
         the single-body part of $H_N$ and $H_{\rm int}$, respectively. 
         The vertexes with undulating line and a short vertical stump 
         represent $\mathbf{T}_1^{(1)}$ and $\mathbf{T}_2^{(1)}$, respectively.
         }
 \label{pdbl_diag}
\end{figure}

The details of the Goldstone diagrams and the corresponding algebraic 
expressions for the PRCC theory with CCSD approximation are discussed in one 
of our previous works \cite{chattopadhyay-12b}. In a subsequent work 
\cite{chattopadhyay-13b} we also described the linearized PRCC (LPRCC)
equations obtained from the approximation 
$   \left [\bar{H}^{\rm DC}_{\rm N},\mathbf{T}^{(1)}\right ] 
\approx \left [H^{\rm DC}_{\rm N},\mathbf{T}^{(1)}\right ]$ and
$\bar{H}_{\rm int} \approx \mathbf{D} + \left[\mathbf{D},T^{(0)}\right ]$,
where $\bar{H}_{\rm int} = \exp (-T^{(0)})H_{\rm int}\exp(T^{(0)})$. 
The eigenvalue equation in the PRCC theory is then
\begin{equation}
  \left [H_{\rm N}^{\rm DCB},\mathbf{T}^{(1)}\right ] |\Phi_0\rangle
      =  \bigg ( \mathbf{D} + \left [\mathbf{D},T^{(0)}\right ]  
     \bigg )|\Phi_0\rangle. \;\;\;
   \label{prcc_eq2}
\end{equation}
The cluster operator equations are as given in ref. \cite{chattopadhyay-13b}.
However, in terms of the cluster amplitudes, the equation for the 
$\mathbf{T}^{(1)}_1$ cluster amplitudes is
\begin{eqnarray}
  & & \mathbf{d}^{p}_{a} + \sum_q \mathbf{d}^{p}_{q} t^q_a  
      - \sum_b \mathbf{d}^{b}_{a} t^p_b + \sum_{bq}\Big ( \mathbf{d}^{b}_{q}
        \tilde t^{qp}_{ba} + \tilde {g}^{bp}_{qa} \bm{\tau}_b^q \Big )
                          \nonumber \\
  & & + \sum_{bqr}{\tilde g}^{bp}_{qr} \bm{\tau}_{ba}^{qr}  
      - \sum_{bcq} g_{qa}^{bc} \tilde{\bm{\tau}}_{bc}^{qp} 
      + \left ( \varepsilon _p - \varepsilon _a \right ) \bm{\tau}_a^p = 0,
  \label{psing_eqn}        
\end{eqnarray}
where $\mathbf{d}^{j}_{i} = \langle j|\mathbf{d}|i\rangle$ is the matrix 
element of the dipole operator, 
$\tilde{g}_{ij}^{kl} = g_{ij}^{kl} - g_{ji}^{kl}\equiv g_{ij}^{kl}-g_{ij}^{lk}$ 
is the antisymmetrized matrix element of the two-body interaction and 
similarly, $\tilde{\tau} $ is the antisymmetrized perturbed cluster 
amplitudes. The Goldstone diagrams arising from the terms in the 
equation are given in Fig. \ref{psing_diag}. Similarly, the LPRCC equation for
the $\mathbf{T}^{(1)}_2$ cluster amplitudes is 
\begin{eqnarray}
  & &   \bigg [ \sum_r \Big ( \mathbf{d}_r^pt_{ab}^{rq} 
      + g_{rb}^{pq}\bm{\tau}_a^r \Big ) - \sum_c \Big ( \mathbf{d}^{c}_{a}
        t_{cb}^{pq} + g_{ab}^{cq} \bm{\tau}_c^p \Big )
      + \sum_{rc} \Big ( g_{ar}^{pc} \tilde{\bm{\tau}}_{cb}^{rq} 
                  \nonumber \\
  & & - g_{rb}^{pc} \bm{\tau}_{ac}^{rq}  - g_{ar}^{cp} \bm{\tau}_{cb}^{rq}  
        \Big ) \bigg ] + \left [ \begin{array}{c}
                                   p\leftrightarrow q \\
                                   a\leftrightarrow b 
                                 \end{array} \right ] 
      + \sum_{rs} g_{rs}^{pq} \bm{\tau}_{ab}^{rs}
      + \sum_{cd} g_{ab}^{cd} \bm{\tau}_{cd}^{pq} 
                  \nonumber \\
  & & + \left ( \varepsilon_p 
                + \varepsilon_q - \varepsilon_a - \varepsilon_b\right )
        \bm{\tau}_{ab}^{pq} = 0,
  \label{pdbl_eqn}                         
\end{eqnarray}
where $\bigl( \begin{smallmatrix}p\leftrightarrow q \\ a\leftrightarrow b
\end{smallmatrix} \bigr )$ represents terms similar to those in
$[\cdots] $  but with the combined permutations $p\leftrightarrow q$ and
$a\leftrightarrow b$. The Goldstone diagrams arising from the terms 
in the above equation are shown in Fig. \ref{pdbl_diag}. However, as 
discussed earlier in the case of LCCSDT, it is more appropriate to write the 
cluster amplitude equations in terms of the reduced matrix elements. 
For this we define the cluster amplitude of $\mathbf{T}_1^{(1)}$ as 
$\bm{\tau}_{a,1}^b$, where the bold face is to indicate that the cluster
amplitude correspond to a rank one operator and subscript `1' is to indicate 
the rank of the operator. 
 As mentioned earlier, the PRCC theory is general and 
applicable to perturbations with operators of any rank in the electron 
sector. So, for other forms of perturbations, the index `1' may be replaced
with the appropriate rank.
This definition effectively subsumes the reduced 
matrix element of the $c$-tensor in the definition of $\mathbf{T}_1^{(1)}$ 
given in Eq. (\ref{pt1_def}). Similarly, cluster amplitude of 
$\mathbf{T}_2^{(1)}$ is
defined as $\bm{\tau}_{ab,l_1l_2}^{pq}$, where $l_1$ and $l_2$ are the 
ranks of the $c$-tensor operators coupled to a rank one operator. With this
definition reduced matrix elements of the $c$-tensor part of the 
representation in Eq. (\ref{pt2_def}) is incorporated to the definition of 
$\bm{\tau}_{ab,l_1l_2}^{pq}$. Following similar procedure as in LCCSDT, the 
linearized PRCC equations of the cluster amplitudes $\bm{\tau}_{a,1}^p$ and 
$\bm{\tau}_{ab,l_1l_2}^{pq}$ in terms of reduced matrix elements are
\begin{widetext}
\begin{eqnarray}
  & &  \mathbf{d}_{a,1}^{p}  
       + \sum_q {\cal A}_1\mathbf{d}_{q,1}^{p}t^q_a              
       - \sum_b {\cal A}_2\mathbf{d}_{a,1}^{b}t^p_b              
       + \sum_{bq}\mathbf{d}_{q,1}^{b}\left ( 
              {\cal A}_3 t_{ba,1}^{qp}                           
              - \sum_{k} {\cal A}_4 t_{ab,k}^{qp} \right )       
       + \sum_{bq} \bm{\tau}_{b,1}^q\left (
              {\cal A}_5 g_{aq,1}^{pb}                           
              - \sum_{k} {\cal A}_6 g_{aq,k}^{bp} \right)        
                             \nonumber \\
  & & + \sum_{bqr}\sum_{m_1 m_2}\bm{\tau}_{ba,m_1m_2}^{qr} 
             \left ( {\cal A}_7 g_{rq,m_2}^{pb}                  
             - \sum_{k} {\cal A}_8  g_{qr,k}^{pb}  \right )      
      - \sum_{bcq} \sum_{m_1 m_2}  \left (
             {\cal A}_9 g_{aq,m_2}^{cb} \bm{\tau}_{cb,m_1m_2}^{pq} 
             - \sum_{k}{\cal A}_{10} g_{aq,k}^{cb}
               \bm{\tau}_{bc,m_1m_2}^{pq}                        
               \right ) 
                             \nonumber \\
  & & + \left ( \tilde{\varepsilon} _p 
               - \tilde{\varepsilon} _a \right ) \bm{\tau}_a^p = 0, 
        \label{lprcc_s}\\
  & &   \Bigg ( 
        \sum_{r}  {\cal B}_1 \mathbf{d}^{p}_{r} t^{ab,l_2}_{rq}  
      - \sum_{c}  {\cal B}_2 \mathbf{d}^{c}_{a} t_{cb,l_2}^{pq}  
      + \sum_{r}  {\cal B}_3 g_{rb,l_2}^{pq} \bm{\tau}^r_a       
      - \sum_{c}  {\cal B}_4 g_{ab,l_2}^{cq}\bm{\tau}^p_c        
      + \sum_{rc} {\cal B}_5 g_{ar,l_1}^{pc}                     
                             \bm{\tau}_{cb,l_1l_2}^{rq} 
      - \sum_{rc} \sum_{m_1m_2} {\cal B}_6 g_{ar,l_1}^{pc}       
                             \bm{\tau}_{bc,m_1m_2}^{rq} 
                                      \nonumber \\
  & & - \sum_{rc} \sum_{km_1m_2} {\cal B}_7 g_{rb,k}^{pc}        
                             \bm{\tau}_{ac,m_1m_2}^{rq}  
      - \sum_{rck} {\cal B}_8 g_{ar,k}^{cp}                      
                             \bm{\tau}_{cb,l_1l_2}^{rq} 
              \Bigg )
      + \Bigg ( \begin{array}{c}
                  p\leftrightarrow q \\
                  a\leftrightarrow b 
                \end{array} \Bigg ) 
      + \sum_{rs}\sum_{km_1m_2} {\cal B}_9 g_{rs,k}^{pq}        
                             \bm{\tau}_{ab,m_1m_2}^{rs}
                                      \nonumber \\
  & & + \sum_{cd} \sum_{km_1m_2}{\cal B}_{10}g_{ab,k}^{cd}      
                             \bm{\tau}_{cd,m_1m_2}^{pq}
      + \left ( \tilde{\varepsilon}_p 
                + \tilde{\varepsilon}_q 
                - \tilde{\varepsilon}_a 
                - \tilde{\varepsilon}_b\right )\bm{\tau}_{ab,l_1l_2}^{pq}= 0,
   \label{lprcc_d}
\end{eqnarray}
\end{widetext}
where ${\cal A} $ and ${\cal B} $ are the angular coefficients listed in the
Appendix \ref{app_d}-\ref{app_e} and 
$\mathbf{d}_{i,1}^{j} = \langle j||\mathbf{d}||i\rangle $ is the 
reduced matrix element of the electric dipole operator. In the above 
equations, unlike in Eqs. (\ref{psing_eqn}) and (\ref{pdbl_eqn}), each of the 
terms are written separately without symmetrization. This is essential as the 
direct and exchange diagrams have different angular factors and summation 
indexes. 

Although we include $T_3^{(0)}$ in the calculations of the unperturbed
cluster equations, in the PRCC theory computations we restrict to single and 
double approximation. The reasons for this are the large number of cluster
amplitudes and a rather involved angular integration for the diagrams 
associated with $\mathbf{T}_3^{(1)}$. We, however, consider the contributions
from approximate $\mathbf{T}_3^{(1)}$ obtained through perturbative 
calculations. For this we consider the dominant perturbative term, and the
details are provided in the next Section. 


\section{Dipole Polarizability}

\subsection{Expression of $\alpha$ in PRCC}

 The electric dipole polarizability of the ground state of a closed-shell
atom is given by
\begin{equation}
  \alpha = -2  \sum_{I} \frac
  {\langle \Psi_0|\mathbf D|\Psi_I\rangle \langle \Psi_I|\mathbf D|
  \Psi_0\rangle}{E_0 - E_I}, 
\end{equation}
where $|\Psi_I \rangle $ are the intermediate atomic states and $E_I$ is the 
energy of the atomic state. Considering that the ground state of a 
closed-shell atom or ion is even parity, $|\Psi_I \rangle $ must be odd
parity states as $\mathbf{D}$ is an odd parity operator. The above expression 
of $\alpha$ in terms of the PRCC theory is
\begin{equation}
  \alpha = -\frac{\langle \Phi_0|\mathbf{T}^{(1)\dagger}\bar{\mathbf{D}} + 
   \bar{\mathbf{D}}\mathbf{T}^{(1)}|\Phi_0\rangle}{\langle\Psi_0|\Psi_0\rangle},
\end{equation}
where, $\bar{\mathbf{D}} = e^{{T}^{(0)\dagger}}\mathbf{D} e^{T^{(0)}}$, 
represents the unitary transformed electric dipole operator and 
$\langle\Psi_0|\Psi_0\rangle$ is the normalization factor. 
Following the derivations presented in our previous works
\cite{chattopadhyay-13a,chattopadhyay-14a}, retaining terms 
up to quadratic in cluster operators, we can write
\begin{eqnarray}
 \alpha & \approx & \frac{1}{\cal N}\langle\Phi_0|
     \mathbf{T}_1^{(1)\dagger}\mathbf{D} + \mathbf{D}\mathbf{T}_1^{(1)} 
     + \mathbf{T}_1^{(1)\dagger}\mathbf{D}T_1^{(0)} 
     + T_1^{(0)\dagger}\mathbf{D}\mathbf{T}_1^{(1)}\nonumber \\
    &&  + \mathbf{T}_2^{(1)\dagger}\mathbf{D}T_1^{(0)} 
     + T_1^{(0)\dagger}\mathbf{D}\mathbf{T}_2^{(1)}
     + \mathbf{T}_1^{(1)\dagger}\mathbf{D}T_2^{(0)}\nonumber \\ 
    && + T_2^{(0)\dagger}\mathbf{D}\mathbf{T}_1^{(1)}
     + \mathbf{T}_2^{(1)\dagger}\mathbf{D}T_2^{(0)} 
     + T_2^{(0)\dagger}\mathbf{D}\mathbf{T}_2^{(1)}
     |\Phi_0\rangle, 
 \label{exp_alpha}
\end{eqnarray}
where ${\cal N} = \langle\Phi_0|\exp[T^{(0)\dagger}]\exp[T^{(0)}]
|\Phi_0\rangle$ is the normalization factor, which involves a non-terminating
series of contractions between ${T^{(0)}}^\dagger $ and $T^{(0)} $. 
In the present work we use 
${\cal N} \approx \langle\Phi_0|T_1^{(0)\dagger}T_1^{(0)} + 
T_2^{(0)\dagger}T_2^{(0)}|\Phi_0\rangle$. It must be mentioned here that,
as discussed in our previous work \cite{chattopadhyay-14a}, the expression
of $\alpha$ involves only connected diagrams and the normalization
factor is essential. From the above expression of $\alpha$,
an evident advantage of calculation using PRCC theory is the absence of
summation over $|\Psi_I\rangle $. The summation is subsumed in the 
evaluation of the $\mathbf{T}^{(1)}$ in a natural way. This is 
one of the key advantage of using PRCC theory. 
\begin{figure}[h]
 \includegraphics[width=8.5cm]{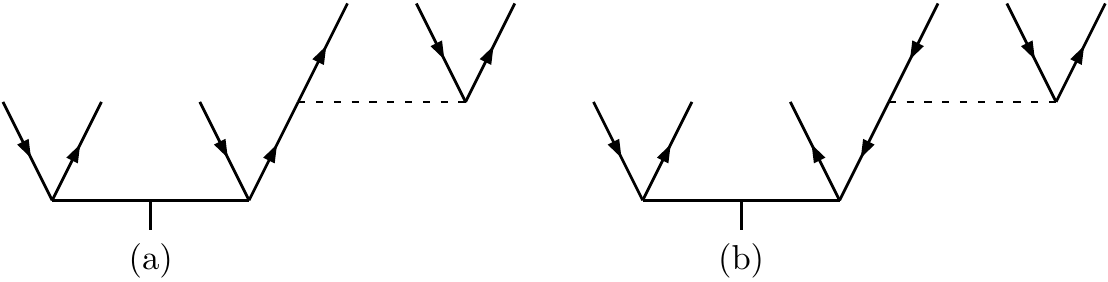}
 \caption{Goldstone diagrams of approximate $\mathbf{T}_3^{(1)}$ obtained 
          from perturbing $\mathbf{T}_2^{(1)}$ with one order of the 
          electron-electron interaction $g=1/r_{12} + g^{\rm B}_{12}$, 
          represented by dashed lines in the diagrams.
         }
 \label{t3_ptr}
\end{figure}


\subsection{Perturbative $\mathbf{T}_3^{(1)}$ and $\alpha$}

\begin{figure}[h]
 \includegraphics[width=8.5cm]{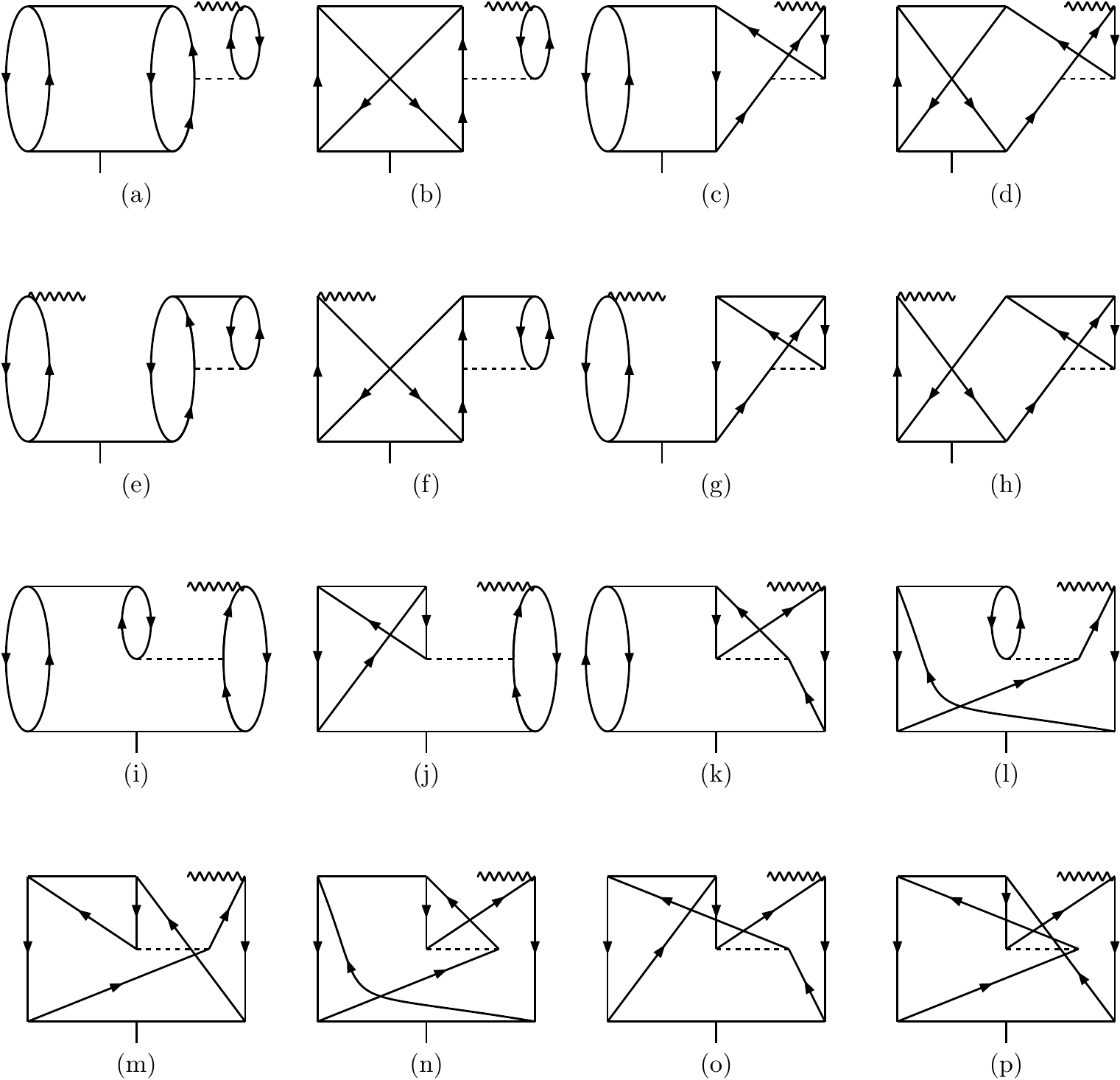}
 \caption{Diagrams of $\alpha$ which arise from,
          $T_2^{(0)\dagger}\mathbf{D}\mathbf{T}_3^{(1)}$, which represents
          the perturbative $\mathbf{T}_3^{(1)}$ contracted with 
          $T_2^{(0)\dagger}$ and $\mathbf{D}$. The $\mathbf{T}_3^{(1)}$, 
          considered in the diagrams, is obtained from particle-particle 
          contraction of $\mathbf{T}_2^{(1)}$ and electron-electron interaction 
          Hamiltonian $g=1/r_{12} + g^{\rm B}_{12}$, represented by dashed 
          lines in the diagrams. 
         }
 \label{t3_ptrpp}
\end{figure}

  To obtain the dominant contributions from the triple excitation cluster 
operators in PRCC, $\mathbf{T}_3^{(1)}$, we consider the perturbative 
approximation. In this scheme $\mathbf{T}_3^{(1)}$ is approximated as a 
first order perturbation to $\mathbf{T}_2^{(1)}$, and it accommodates the 
leading order terms in the cluster amplitude equations of $\mathbf{T}_3^{(1)}$.
There are two diagrams in this approximation and are shown in 
Fig. \ref{t3_ptr}, and these combine to give the perturbative triple
excitation cluster amplitude
\begin{equation}
 \bm{\tau}_{abc}^{pqr} \approx \frac{1}{\Delta\epsilon_{pqr}^{abc} } 
        \Big ( \sum_s\bm{\tau}_{ab}^{ps}g^{qr}_{sc} 
               - \sum_d \bm{\tau}_{ad}^{pq}g^{dr}_{bc} \Big ),
\end{equation}
where, $\Delta\epsilon_{pqr}^{abc}= \epsilon_p + \epsilon_q + \epsilon_r -
\epsilon_a - \epsilon_b - \epsilon_c$, and as defined earlier 
$g_{ij}^{kl} = \langle kl| 1/r_{12} + g^{\rm B}_{12}|ij\rangle $. The first 
and second term on the right hand side of the above equation correspond to the 
Goldstone diagrams in Fig. \ref{t3_ptr}(a) and (b), respectively. Each of these
diagrams, after contraction with $T_2^{(0)\dagger}$ and $\mathbf{D}$,
generate sixteen diagrams of $\alpha$ each. For example, the set of the
sixteen diagrams arising from the perturbative $\mathbf{T}_3^{(1)}$ 
represented by Fig. \ref{t3_ptr}(a) are shown in Fig. \ref{t3_ptrpp}.
The other term associated with $\mathbf{T}_3^{(1)}$ which contributes to 
$\alpha$  is $T_1^{(0)\dagger} T_1^{(0)\dagger}\mathbf{D}\mathbf{T}_3^{(1)}$.
We, however, neglect this as it is second order in $T_1^{(0)\dagger}$ and 
expect the contribution to be smaller than 
$T_1^{(0)\dagger}\mathbf{D}\mathbf{T}_2^{(1)}$, which as we shall discuss
later has the smallest contribution in the expression of $\alpha$ 
in Eq. (\ref{exp_alpha}). Here after, the values of $\alpha$ obtained with 
the inclusion of perturbative $\mathbf{T}_3^{(1)}$ are referred to as 
PRCC(T).


\section{Computational details}


\subsection{Basis set and nuclear density}

 We use Gaussian type orbitals (GTOs) \cite{mohanty-91}, and the details 
relevant to the use of GTOs in RCC and PRCC are described in our previous
works \cite{chattopadhyay-12a, chattopadhyay-13a}. The GTOs are finite basis 
set orbitals and are the linear combinations of Gaussian type functions (GTFs).
The exponents of the GTFs are defined in terms of two parameters $\alpha_{0}$ 
and $\beta$. We
consider even tempered basis set, or in other words, different $\alpha_{0}$ 
and $\beta$ for orbitals of each $j$. We also use kinetic balance condition
\cite{stanton-84} to obtain small components of the orbitals from the 
large component. Further more, it is appropriate to incorporate Breit 
interaction \cite{quiney-91} in the generation of GTOs as the 
present study includes Hg, a high $Z$ atom. For this the works of 
Quiney \cite{quiney-03} and Mohanty \cite{mohanty-91}, and their collaborators 
are excellent references. Keeping in view the implementations general and
incorporating mathematically intricate interaction Hamiltonians, for example,
the Uehling potential, we generate the GTOs on a grid \cite{chaudhuri-99} 
with $V^{\rm N}$ potential. The basis parameters $\alpha_{0}$ and $\beta$
are optimized by matching the orbital and self-consistent field (SCF)
energies obtained from GRASP2K \cite{jonsson-13} with the Dirac-Coulomb
Hamiltonian. The values of the optimized parameters of Zn, Cd and Hg are 
listed in Table. \ref{basis}.
\begin{table}[h]
   \caption{The $\alpha_0$ and $\beta$ parameters for the $s$, $p$
            and $d$ orbitals of the even tempered GTO basis used in the 
            present calculations.}
   \label{basis}
   \begin{tabular}{cccccccc}
   \hline
   \hline
     Atom & \multicolumn{2}{c}{$s$} & \multicolumn{2}{c}{$p$} & 
     \multicolumn{2}{c}{$d$}  \\
     & $\alpha_{0}$  & $\beta$ & $\alpha_{0}$ & $\beta$  
     & $\alpha_{0}$  & $\beta$  \\
     \hline
     Zn &\, 0.0385  &\, 2.045 &\, 0.1095  &\, 2.035 &\, 0.0091  &\, 2.010 \\
     Cd &\, 0.0505  &\, 2.101 &\, 0.0775  &\, 1.985 &\, 0.0340  &\, 1.950 \\
     Hg &\, 0.0505  &\, 2.045 &\, 0.1019  &\, 2.223 &\, 0.0380  &\, 2.050 \\
     \hline
   \end{tabular}
\end{table}

  The SCF energies $E_{\rm SCF}$ obtained with the optimized basis 
parameters are listed in Table. \ref{scf}. It is evident from 
the table that $E_{\rm SCF}$ from the GTOs are in very good agreement with 
the results of GRASP2K, which solves the Dirac-Hartree-Fock equations 
numerically. An important step in generating the orbitals with
GRASP2K is, we use the Hartree-Fock orbitals \cite{fischer-87} as the 
starting values of GRASP2K to improve convergence. As mentioned earlier,
we also compare the orbital energies for basis parameter optimization.
The details of these comparisons are presented and described in the 
results and discussions section.
\begin{table}[h]
        \caption{The Dirac-Coulomb SCF energies $E_{\rm SCF}$ of Zn, Cd and Hg 
                 obtained from GRASP2K \cite{jonsson-13} and using Gaussian 
                 type orbitals are listed. The Breit interaction corrections 
                 to SCF energy $\Delta{E}^{\rm SCF}_{\rm Br}$ are computed 
                 using the Gaussian type orbitals. All the values are in 
                 atomic units (hartree).
                }
        \label{scf}
        \begin{center}
        \begin{tabular}{ldddd}
            \hline
            Atom &
              \multicolumn{2}{c}{$E_{\rm SCF}$} &
              \multicolumn{2}{c}{$\Delta{E}^{\rm SCF}_{\rm Br}$} \\
            \hline
             &\multicolumn{1}{c}{GTO} &
              \multicolumn{1}{c}{GRASP2K}  &
              \multicolumn{1}{c}{Present}  &
              \multicolumn{1}{c}{Ref. \cite{ishikawa-94}} \\
            \hline
            Zn & -1794.6127   & -1794.6127   & -0.7610  & -0.7610  \\
            Cd & -5593.3188   & -5593.3184   & -3.8389  & -3.8389  \\
            Hg & -19648.8243  & -19648.8580  & -22.6328 & -22.6325 \\
           \hline
        \end{tabular}
        \end{center}
\end{table}

To generate the nuclear potential $V_N(r)$, we use two-parameter
finite size Fermi density distribution of the nucleus
\begin{equation}
   \rho_{\rm nuc}(r) = \frac{\rho_0}{1 + e^{(r-c)/a}},
\end{equation}
where, $a=t 4\ln(3)$. The parameter $c$ is the half charge radius so that 
$\rho_{\rm nuc}(c) = {\rho_0}/{2}$ and $t$ is the skin thickness. 
Using the orbital basis set, we can then solve the RCC and PRCC equations
with standard linear algebra method. For efficient parallel implementation
we solve the equations iteratively using Jacobi method. It is, however,
a method with slow convergence, so employ direct inversion in the 
iterated subspace (DIIS) \cite{pulay-80} to improve convergence.


\subsection{Breit and vacuum polarization corrections}

  In the present work,  we use the general expressions of Breit interaction
integrals listed in the work of Grant \cite{grant-76}. To examine the 
corrections to orbital energies arising from the Breit interactions, we 
generate the orbitals as solutions of two slightly different
single particle equations. In the first case, the orbitals $|\psi_i \rangle$
are computed with the Dirac-Hartree-Fock (DHF) potential and solutions
of the equation,
\begin{equation}
   \left ( h_{0} + U_{\rm DHF} \right ) |\psi_i \rangle = 
   \epsilon_i |\psi_i \rangle,  \nonumber  
\end{equation}
where, 
$h_0 = c\bm{\alpha}\cdot\mathbf{p} +(\beta - 1)c^2 -V_{\rm N}(\mathbf{r})$ is
the single particle part of Dirac-Coulomb Hamiltonian, $|\psi_i \rangle$ is a 
four component orbital and $\epsilon_i$ is the corresponding eigenvalue.
The DHF potential in the above equation is defined as
\begin{equation}
  U_{\rm DHF}|\psi_i \rangle = \sum_c^{\rm core} \left [
    \langle \psi_c|\frac{1}{r_{12}}\left (1-P_{12} \right )|\psi_c\rangle 
    |\psi_i\rangle \right ],
  \label{dhf_pot}
\end{equation}
where, $P_{12}$ is the permutation operator to represent the exchange
integral, $c$ represents core orbitals and `core' indicates sum over all
the core orbitals. This implies that the core orbitals are solutions of 
a set of coupled integro-differential equations and solved using 
self-consistent-field (SCF) methods. In the second case, we compute 
the orbitals $|\psi'_i \rangle$ with the Dirac-Hartree-Fock-Breit (DHFB)
potential. The orbitals are then the solutions of the single particle
equation
\begin{equation}
   \left ( h_{0}  + U_{\rm DHFB} \right) |\psi'_i \rangle = 
   \epsilon'_i |\psi '_i \rangle,  \nonumber  
\end{equation}
where, $\epsilon'_i$ is the eigenvalue with the DHFB potential, and 
$U_{\rm DHFB}$ is obtained by adding $g^{\rm B}_{12}$ to the central 
potential in Eq. (\ref{dhf_pot}). From the solutions we define the 
correction to orbital energies due to Breit interaction as
\begin{equation}
   \Delta\epsilon_{{\rm Br}(i)} = \epsilon' _i - \epsilon_i.
\end{equation}
In a similar way, we also compute the correction due to Uehling potential
$\Delta\epsilon_{\rm Ueh}$. 

  From the two sets of the orbitals, we define two many-particle ground 
state reference $|\Phi_0\rangle$ and $|\Phi'_0\rangle$, which 
are determinantal states consisting of $|\psi_c \rangle$ and 
$|\psi'_c \rangle$ orbitals, respectively. Based on these states, the
SCF energy correction due to Breit interaction is
\begin{equation}
  \Delta E_{\rm Br}^{\rm SCF} = \langle\Phi'_0|H^{\rm DCB}|\Phi'_0\rangle
        - \langle\Phi_0|H^{\rm DC}|\Phi_0\rangle,
\end{equation}
where, $H^{\rm DC}$ is the Dirac-Coulomb Hamiltonian: the Hamiltonian
$H^{\rm DCB}$ defined in Eq. (\ref{ham_dcb}) without the Breit interaction.
The values of $\Delta E_{\rm Br}^{\rm SCF}$ for Zn, Cd and Hg are listed in
Table. \ref{scf}, and are near perfect match with the values reported in 
a previous work \cite{ishikawa-94}. This is another important comparison 
which validates the choice of the optimized basis set parameters used in the
present study. In the results and discussions section, we present
$\Delta\epsilon_{\rm Br}$  of Zn, Cd and Hg orbitals. For the first two atoms,
Zn and Cd, we were unable to get previous results from the literature. 
However, for Hg a previous work \cite{lindroth-89a} has provided the values of
$\Delta\epsilon_{\rm Br}$, and our results are in excellent agreement with
those values.

  Another way to quantify the effect of Breit interaction is to calculate
the first order correction to the SCF energy as
\begin{equation}
  \langle H^{\rm B}\rangle_{\rm DF} = \langle\Phi_0|\sum_{i<j}g^{\rm B}(r_{ij})
       |\Phi_0\rangle.
\end{equation}
In a previous work \cite{chattopadhyay-12b}, we have reported 
$\langle H^{\rm B}\rangle_{\rm DF}$ for the noble gas atoms, and computations
were based on the compact expressions of Breit interaction integrals listed in 
the work of Grant and McKenzie \cite{grant-80}. The computation of
$\langle H^{\rm B}\rangle_{\rm DF}$ is well suited for testing the 
implementation of Breit interactions. In the present work, as we have 
incorporated Breit interaction in the GTO generation and coupled-cluster 
codes, we give our results of $\Delta E_{\rm Br}^{\rm SCF}$ and
$\Delta\epsilon_{\rm Br}$, but not the values of
$\langle H^{\rm B}\rangle_{\rm DF}$. It must be mentioned here that,  among 
the previous works on Breit interactions, there is another approach to 
evaluate the Breit interaction matrix elements reported in the work of Mann 
and Johnson \cite{mann-71}. It is based on the coupling of the Dirac 
matrices with the angular part of the orbitals. In contrast, the expressions
of Grant and collaborators, which we have used, are based on the expansion
of $g^{\rm B}(r_{12})$ as linear combination of irreducible tensor operators.


\section{Results and Discussions}

  The elements of the group IIB studied in the present work, have filled
$ns$ orbitals as valence shells and in this regard, similar to the neutral 
alkaline-earth-metal atoms. There is, however, an important difference:
in the group IIB elements the filled $(n-1)d$ shells are the highest energy 
core orbitals and we can expect significant contribution to the correlation 
effects from the electrons in the $(n-1)d$ shell. This is indeed the case
and is reflected in the identification of the occupied orbitals with 
dominant contributions to the leading order (LO) term, 
$\mathbf{T}_1^{(1)\dagger}\mathbf{D} + \text{H.c.}$, 
in $\alpha$. We also examine the trends in the contribution from 
Breit-interaction to the energies of the occupied orbitals.  For better 
description the results for each of the elements (Zn, Cd and Hg) are 
discussed separately. All the values of $\alpha$ are in atomic units, that 
is in units of $a_0^3$, where $a_0$ is the Bohr radius.

\begin{table}[h]
        \caption{Orbital energies of Zn and Cd obtained from GRASP2K 
                 \cite{jonsson-13} and Gaussian type orbitals in atomic units
                 (hartree). Here [x] represents multiplication by ${10^x}$.}
        \label{orbe_zncd}
        \begin{tabular}{lddD..{3.7}D..{3.7}}
            \hline
              {Orbital} 
              & \multicolumn{1}{c}{\textrm{GRASP2K}}     & 
                \multicolumn{1}{c}{\textrm{DC}}          &
                 \multicolumn{1}{c}{\text{$\Delta\epsilon_{\rm Br}$}}         &
                 \multicolumn{1}{c}{\text{$\Delta\epsilon_{\rm Ueh}$}}   \\
            \hline
              \multicolumn{5}{c}{Zn}\\
            $1s_{1/2}$ & -357.7486 & -357.7486 & 4.364[-1]  &-2.174[-2] \\ 
            $2s_{1/2}$ & -45.3461  & -45.3461  & 3.129[-2]  &-2.125[-3] \\
            $2p_{1/2}$ & -39.7403  & -39.7402  & 5.524[-2]  & 1.724[-4] \\
            $2p_{3/2}$ & -38.8513  & -38.8513  & 3.586[-2]  & 1.859[-4] \\
            $3s_{1/2}$ & -5.8000   & -5.7999   & 3.301[-3]  &-3.113[-4] \\
            $3p_{1/2}$ & -3.9579   & -3.9578   & 5.896[-3]  & 3.201[-5] \\
            $3p_{3/2}$ & -3.8372   & -3.8371   & 3.135[-3]  & 3.419[-5] \\  
            $3d_{3/2}$ & -0.7709   & -0.7709   & 2.015[-4]  & 2.470[-5] \\ 
            $3d_{5/2}$ & -0.7547   & -0.7547   &-8.255[-4]  & 2.453[-5] \\
            $4s_{1/2}$ & -0.2986   & -0.2986   & 1.251[-4]  &-1.080[-5] \\
              \multicolumn{5}{c}{Cd}\\
            $1s_{1/2}$ & -987.3591 & -987.3580 & 2.017      &-1.519[-1] \\
            $2s_{1/2}$ & -149.8044 & -149.8032 & 1.810[-1]  &-1.702[-2] \\
            $2p_{1/2}$ & -139.0231 & -139.0218 & 3.117[-1]  & 5.952[-4] \\
            $2p_{3/2}$ & -131.9158 & -131.9145 & 2.109[-1]  & 1.006[-3] \\
            $3s_{1/2}$ & -29.3222  & -29.3212  & 2.465[-2]  &-3.239[-3] \\
            $3p_{1/2}$ & -24.9552  & -24.9541  & 4.581[-2]  & 1.674[-4] \\
            $3p_{3/2}$ & -23.6459  & -23.6451  & 2.743[-2]  & 2.552[-4] \\
            $3d_{3/2}$ & -16.0009  & -16.0001  & 1.231[-2]  & 2.389[-4] \\
            $3d_{5/2}$ & -15.7383  & -15.7374  & 4.173[-3]  & 2.344[-4] \\
            $4s_{1/2}$ & -4.7469   & -4.7460   & 3.487[-3]  &-5.810[-4] \\
            $4p_{1/2}$ & -3.2707   & -3.2698   & 6.390[-3]  & 5.666[-5] \\
            $4p_{3/2}$ & -3.0461   & -3.0451   & 3.141[-3]  & 7.209[-5] \\
            $4d_{3/2}$ & -0.7383   & -0.7374   & 8.961[-5]  & 5.591[-5] \\
            $4d_{5/2}$ & -0.7089   & -0.7080   &-8.777[-4]  & 5.506[-5] \\
            $5s_{1/2}$ & -0.2814   & -0.2810   & 1.930[-4]  &-3.273[-5] \\
           \hline
        \end{tabular}
\end{table}


\subsection{Zn}

  The corrections to the orbitals energies $\Delta\epsilon_{\rm Br}$
and $\Delta\epsilon_{\rm Ue}$ arising from Breit-interaction and 
Uehling potential, respectively, are listed in Table. \ref{orbe_zncd}. From the
table it is evident that the Breit-interaction tends to {\em relax} the 
orbitals as $\Delta\epsilon_{\rm Br}$ is positive in all the cases except
$3d_{5/2}$. For the latter, $\Delta\epsilon_{{\rm Br}(3d_{5/2})}$, is negative 
and indicates contraction of the orbital. In absolute terms the value of 
$-8.255\times 10^{-4}$ hartree for $\Delta\epsilon_{{\rm Br}(3d_{5/2})}$ is 
small but the magnitude is larger than $\Delta\epsilon_{{\rm Br}(3d_{3/2})}$. 
As to be expected, the deeper core orbitals or orbitals with lower principal 
quantum number $n$ have larger $\Delta\epsilon_{\rm Br}$ and there is a three 
orders of magnitude difference between the values of $\Delta\epsilon_{\rm Br}$ 
for $1s$ and $4s$.
\begin{table}[h]
  \caption{Convergence pattern of $\alpha$ for Zn and Cd as function of
           the basis set size. The values of $\alpha$ are in atomic units
           ( $a_0^3$).}
  \label{conv_pat}
  \begin{tabular}{lcc}
      \hline
      No. of orbitals & Basis size & $\alpha $   \\
      \hline
              \multicolumn{3}{c}{Zn}\\
      113 & $(15s, 13p, 11d, 9f, 9g, 7h)    $ & 38.722  \\
      135 & $(17s, 15p, 15d, 10f, 10g, 9h)  $ & 38.717  \\
      153 & $(19s, 17p, 17d, 11f, 11g, 11h) $ & 38.716  \\
      171 & $(21s, 19p, 19d, 13f, 13g, 11h) $ & 38.716  \\
              \multicolumn{3}{c}{Cd}\\
      99  & $(15s, 12p, 11d,  7f,  6g,  6h) $ & 49.421  \\
      121 & $(17s, 14p, 13d,  9f,  8g,  8h) $ & 49.135  \\
      143 & $(19s, 16p, 15d, 11f, 10g, 10h) $ & 49.113  \\
      165 & $(21s, 18p, 17d, 13f, 12g, 12h) $ & 49.112  \\
              \multicolumn{3}{c}{Hg}\\
      112 & $(12s, 11p, 11d, 11f,  9g,  8h) $ & 33.513  \\
      134 & $(14s, 13p, 13d, 13f, 11g, 10h) $ & 33.499  \\
      167 & $(17s, 16p, 16d, 16f, 14g, 13h) $ & 33.499  \\
      178 & $(18s, 17p, 17d, 17f, 15g, 14h) $ & 33.499  \\
      \hline
  \end{tabular}
\end{table}

The energy correction arising from the Uehling potential 
$\Delta\epsilon_{\rm Ue} $ are also listed in Table. \ref{orbe_zncd}.  It
is evident that Uehling potential tends to contract the $s$ orbitals 
as $\Delta\epsilon_{\rm Ue}$ of these orbitals are negative. On the other 
hand, the occupied orbitals of other symmetries ($p$ and $d$) {\em relax} 
and are indicated by the positive values of $\Delta\epsilon_{\rm Ue} $. This
trend is similar to the results of doubly ionized alkaline-earth-metals
Mg$^{2+}$, Ca$^{2+}$, Sr$^{2+}$, and Ba$^{2+}$ reported in our previous 
work \cite{chattopadhyay-13b}. In terms of magnitude, the values of 
$\Delta\epsilon_{\rm Ue} $ are on average an order of magnitude smaller than
$\Delta\epsilon_{\rm Br}$.

 From Table. \ref{orbe_zncd}, it is evident that the basis set parameters 
reproduces the numerical values of the orbital energies, obtained using
GRASP2K \cite{jonsson-13}, to an accuracy of $10^{-4}$ hatree or lower. 
To determine the optimal orbital basis set, we compute $\alpha$ with 
increasing basis size and the results are listed in Table. \ref{conv_pat}. 
From the table, we observe convergence of $\alpha$ up to $10^{-3}$ a.u. with 
a basis set of 171 orbitals. Based on the results, we choose the set with 
135 orbitals as the optimal one and use it for more detailed studies. 

 In Table. \ref{pol_group2b} the converged values of $\alpha$ along with the 
previous theoretical results and experimental data are listed for comparison.
From the table it is evident that our result of 38.72 is in very good agreement
with the experimental value of 38.8(8). Among the previous theoretical 
results,  the results from configuration interaction with a semi-empirical
core-polarization potential (CICP) \cite{ye-PRA-08} is on the lower side. 
There are two other theoretical results based on coupled-cluster theory. 
The first \cite{goebel-96} is using non-relativistic Hamiltonian with
finite field approach, where as the second \cite{yashpal-14} uses
Dirac-Coulomb Hamiltonian with the external electric field treated as a 
perturbation. In both the works, the contributions from triple excitations
are included perturbatively. Compared to the experimental value, the results 
from the first work \cite{goebel-96} is on the higher side, but the 
result from the second work \cite{yashpal-14} is close to the experimental 
value. The method used in ref. \cite{yashpal-14} is similar, in the way
the external field is treated as a perturbation and computation of a second set
of cluster amplitudes, to PRCC. However, our result is in 
better agreement with the experimental value. This may be on account of two 
important factors: inclusion of Breit-interaction in the atomic Hamiltonian
and computation of $T_3^{(0)}$ without perturbative approximations. 
With the inclusion of perturbative $\mathbf{T}_3^{(1)}$, 
result listed as PRCC(T) in Table.\ref{pol_group2b}, our
result is in excellent agreement with the experimental data.

 The term wise contribution to $\alpha$ in Eq. (\ref{exp_alpha}) are listed
in Table. \ref{cont_term}. From the table the LO contribution arises from 
$\mathbf{T}_1^{(1)\dagger}\mathbf{D} + \text{H.c.}$ and is larger than
the total value of $\alpha$. This is, perhaps, not surprising as the LO term
subsumes the Dirac-Hartree-Fock contribution and core-polarization effects.
The next to leading order (NLO) is 
$\mathbf{T}_1^{(1)\dagger}\mathbf{D}T_1^{(0)} + \text{H.c.}$,
and opposite in phase to the LO. A similar phase relation between the LO and 
NLO was observed in our previous work on noble gas \cite{chattopadhyay-12b}
and alkaline-Earth-metal \cite{chattopadhyay-14a} atoms. Among the remaining
terms, the contribution from 
$\mathbf{T}_1^{(1)\dagger}\mathbf{D}T_2^{(0)} + \text{H.c.}$ is similar 
in value and phase to the NLO term.
 The sub-shell wise contributions from the LO term, as mentioned earlier is 
the sum of $\mathbf{T}_1^{(1)\dagger}\mathbf{D}$ and its hermitian 
conjugate, are listed in Table. \ref{result_t1d}. From the table, the
valence sub-shell $4s_{1/2}$ is the most dominant, and followed by
$3d_{5/2}$. Both the sub-shell contributions have same phase, and together
accounts for more than 99\% of the LO term.

\begin{table}[h]
  \caption{Static dipole polarizability $\alpha$ of Zn, Cd and Hg 
           in atomic units ($a_0^3$). }
  \label{pol_group2b}
  \begin{center}
  \begin{tabular}{ldcdc}
    \hline
    \multicolumn{1}{c}{Atom}      & \multicolumn{1}{c}{Present}      &
    \multicolumn{1}{c}{Method}    & \multicolumn{1}{c}{Previous Works} &
    \multicolumn{1}{c}{Method}                              \\
    \hline
    \hline
    $\rm{Zn}$ & 38.72  &  PRCC    & 38.12       \text{\cite{ye-PRA-08}}
              & CICP     \\
              & 38.76  &  PRCC(T) & 38.5        \text{\cite{rosenkrantz-80}}
              & MCSCF    \\
              &        &          & 38.4        \text{\cite{roos-05}}
              & CASPT2   \\
              &        &          & 37.86       \text{\cite{kello-95}}
              & CCSD(T)  \\
              &        &          & 38.01       \text{\cite{seth-97}}
              & CCSD(T)  \\
              &        &          & 39.2(8)     \text{\cite{goebel-96}}
              & CCSD(T)  \\
              &        &          & 38.666(35)  \text{\cite{yashpal-14}}
              & RCCSD$_p$T \\
              &        &          & 38.8(8)     \text{\cite{goebel-96}}
              & Expt.    \\
              &        &          & 38.92       \footnotemark[1] & Expt.    \\
    $\rm{Cd}$ & 49.11  &  PRCC    & 44.63       \text{\cite{ye-PRA-08}}
              & CICP     \\
              & 49.20  &  PRCC(T) & 46.9        \text{\cite{roos-05}}
              & CASPT2   \\
              &        &          & 47.63       \text{\cite{kello-95}}
              & CCSD(T)  \\
              &        &          & 46.25       \text{\cite{seth-97}}
              & CCSD(T)  \\
              &        &          & 45.856(42)  \text{\cite{yashpal-14}}
              & RCCSD$_p$T \\
              &        &          & 49.65(1.47) \text{\cite{goebel-95}}
              & Expt.    \\
              &        &          & 49.50       \footnotemark[1] & Expt.    \\
              &        &          & 50.0(2.8)   \footnotemark[2] & Expt.    \\
    $\rm{Hg}$ & 33.50  &  PRCC    & 31.32       \text{\cite{ye-PRA-08}}
              & CICP     \\
              & 33.59  &  PRCC(T) & 33.3        \text{\cite{roos-05}}
              & CASPT2    \\
              &        &          & 33.44       \text{\cite{schwerdtfeger-94}}
              & QCISD(T)  \\
              &        &          & 31.82       \text{\cite{kello-95}}
              & CCSD(T)  \\
              &        &          & 34.42       \text{\cite{seth-97}}
              & CCSD(T)  \\
              &        &          & 34.15       \text{\cite{pershina-08}}
              & CCSD(T)  \\
              &        &          & 33.6        \text{\cite{hachisu-08}}
              & CI + MBPT\\
              &        &          & 33.7(1.3)   \footnotemark[3] & Expt.    \\
              &        &          & 33.75       \footnotemark[4] & Expt.    \\
              &        &          & 33.91(34)   \text{\cite{goebel-96a}}
              & Expt.    \\
    \hline
    \hline
  \end{tabular}
  \end{center}
\footnotetext[1]{Reference \cite{qiao-12} based on experimental data
                 in Ref. \cite{goebel-95, goebel-96}.}
\footnotetext[2]{Reference \cite{goebel-95} based on the refractive
                 index data in Ref. \cite{cuthbertson-1908}.}
\footnotetext[3]{Reference \cite{goebel-96a} based on the dielectric 
                 data in Ref. \cite{wusthoff-36}.}
\footnotetext[4]{Reference \cite{tang-08} based on the experimental data
                  in Ref. \cite{goebel-96a}.}
\end{table}


\subsection{Cd}

  The corrections to the orbitals energies $\Delta\epsilon_{\rm Br}$
and $\Delta\epsilon_{\rm Ue}$ arising from Breit-interaction and 
Uehling potential, respectively, are listed in Table. \ref{orbe_zncd}. 
From the table it is evident that like in Zn $\Delta\epsilon_{\rm Br}$ of 
the $4d_{5/2}$, the sub-shell next to the valence, is negative. Over all the 
general trend in the corrections is very similar to the case of Zn, except 
that the magnitude of the corrections are one order higher. There is,
however, one noticeable change in the relative values of 
$\Delta\epsilon_{\rm Ue}$  for the $p_{1/2}$ and $p_{3/2}$ orbitals. In 
the case of Zn, $\Delta\epsilon_{{\rm Ue}(mp_{1/2})}
\approx \Delta\epsilon_{{\rm Ue}(mp_{3/2})}$ (with $m=2,3$), but in the Cd,
$\Delta\epsilon_{{\rm Ue}(mp_{1/2})}$ is about a factor of two smaller than
$\Delta\epsilon_{{\rm Ue}(mp_{3/2})}$. This indicates an enhanced effect of
the Uehling potential or vacuum polarization potential to the inner $p_{1/2}$
orbitals with higher nuclear charge $Z$. It is an expected trend as the 
$p_{1/2}$ orbitals contract with higher $Z$ due to larger relativistic 
corrections, and the inner orbitals contract more as the correction is 
larger. 
\begin{table}[h]
    \caption{Contribution to $\alpha $ from different terms and their
             hermitian conjugates in the PRCC theory in atomic units
             ($a_0^3$).}
    \label{cont_term}
    \begin{center}
    \begin{tabular}{lddd}
        \hline
        Terms + h.c. & \multicolumn{1}{r}{$\rm{Zn}$}
        & \multicolumn{1}{r}{$\rm{Cd}$}
        & \multicolumn{1}{r}{$\rm{Hg}$}  \\
        \hline
        $\mathbf{T}_1^{(1)\dagger}\mathbf{D} $
        & 45.590   &  61.456   &    41.927     \\
        $\mathbf{T}_1{^{(1)\dagger}}\mathbf{D}T_2^{(0)} $
        & -1.850   & -3.128    & -2.724      \\
        $\mathbf{T}_2{^{(1)\dagger}}\mathbf{D}T_2^{(0)} $
        &  1.364   &  2.060    &  1.504      \\
        $\mathbf{T}_1{^{(1)\dagger}}\mathbf{D}T_1^{(0)} $
        & -1.901   & -3.808    & -1.583      \\
        $\mathbf{T}_2{^{(1)\dagger}}\mathbf{D}T_1^{(0)} $
        &  0.081   &  0.243    &   0.091      \\
        Normalization & 1.118  & 1.157   & 1.171   \\
        Total         & 38.716 & 49.112  & 33.499 \\
        \hline
    \end{tabular}
    \end{center}
\end{table}

 Like in the case of Zn, orbital energies of Cd corresponding to the 
GTOs and numerical results from GRASP2K \cite{jonsson-13} are listed in 
the Table. \ref{orbe_zncd}. It is evident that the basis parameters chosen
for the Cd basis matches the orbital energies with the numerical results to 
within $10^{-4} - 10^{-3}$ hatrees. On comparison,  on an average the 
agreement is in the case of Zn an order of magnitude better. This is on 
account of the larger number of occupied orbitals Cd, which increases the
parameters of optimization. Coming to the results of $\alpha$, from 
Table. \ref{conv_pat}, we find that $\alpha$ converges to 
$\approx 10^{-3}$ a. u. with a basis set of 165 orbitals. However, considering
the number of cluster amplitudes, we take the basis set consisting of
143 orbitals for further computations. It must be mentioned that, with this
basis set the convergence of $\alpha$  is $\approx 10^{-2}$ a. u..

  From the results listed in Table.  \ref{pol_group2b}, it is evident 
that there is a variation in the previous results from coupled-cluster theory.
There are three previous theoretical works on the computation of $\alpha$
using coupled-cluster theory \cite{kello-95,seth-97,yashpal-14}. However,
each of these use different types of basis sets, Ref. \cite{kello-95} and 
\cite{seth-97} are based on optimization with polarization potential and 
pseudo-potential, respectively. In terms of the theory and type of basis 
functions, the methods we have used in the present work is very similar to 
Ref. \cite{yashpal-14}.  There is, however, noticeable difference between
the two results, and this may be due to difference in the methods at various
stages of computations. For the present work, as described earlier, we have 
provided detailed information about the basis set parameters, and convergence 
of $\alpha$ with the basis size.  It must be emphasized that our 
result for $\alpha$ is closest to the experimental value. 
The agreement with the experimental data improves with the 
inclusion of perturbative $\mathbf{T}_3^{(1)}$, the result listed as 
PRCC(T) in Table. \ref{pol_group2b}. 
 The term wise contribution to $\alpha$ as listed in Table. \ref{cont_term} 
has the same trend, albeit larger values, as in Zn.
Coming to the sub-shell contributions to 
the LO term, from the values listed in Table. \ref{result_t1d} the 
pattern is similar to Zn: the dominant contribution arises from the valence 
sub-shell $5s_{1/2}$, and followed by $4d_{5/2}$, the sub-shell below the 
valence. However, compared to Zn, the dominant and next contribution in Cd 
are $\approx 25$ \%  and $\approx 47$\% larger, respectively. 
\begin{table}[h]
    \caption{Four leading contributions to
        $\{ \mathbf{T}_1^{(1)\dagger}\mathbf{D}  \}$ to $\alpha $
        in terms of the core spin-orbitals in atomic units ($a_0^3$). }
    \label{result_t1d}
    \begin{center}
    \begin{tabular}{rrr}
        \hline
          \multicolumn{1}{c}{Zn} & \multicolumn{1}{c}{Cd}
        & \multicolumn{1}{c}{Hg}  \\ \hline
        22.244 (4$s_{1/2}$) & 29.771 (5$s_{1/2}$)  &  17.768 (6$s_{1/2}$) \\
         0.362 (3$d_{5/2}$) &  0.678 (4$d_{5/2}$)  &   2.239 (5$d_{5/2}$) \\
         0.193 (3$d_{3/2}$) &  0.340 (4$d_{3/2}$)  &   0.965 (5$d_{3/2}$) \\
        -0.001 (3$s_{1/2}$) & -0.004 (4$p_{3/2}$)  &  -0.009 (5$p_{3/2}$) \\
        \hline
    \end{tabular}
    \end{center}
\end{table}

 Concerning the experimental results, there is slight variation of the
experimental uncertainty  listed in the literature. In the original 
experimental work of Goebel and Hohm \cite{goebel-95}, the $\alpha$ of Cd is 
reported as $49.65\pm1.46\pm0.16$ a.u. Based on this result the experimental
value is listed as $49.65\pm1.46$,  $49.65 (1.49)$ and $49.65\pm1.62$ in 
Ref. \cite{schwerdtfeger-06}, \cite{mitroy-10}, and \cite{hohm-12}, 
respectively. However, the quadrature of the uncertainties
reported in Ref. \cite{goebel-95} gives the result $49.65(1.47)$, the value
listed in Table. \ref{pol_group2b} of the present work. This is a minor issue 
and does not impact on the experimental results. We have mentioned this to 
explain  the difference in the experimental result of Cd listed in 
Table. \ref{pol_group2b} from the previous works, namely 
Ref. \cite{schwerdtfeger-06}, \cite{mitroy-10} and \cite{hohm-12}. 

One issue which require some consideration is the consistent lower values of 
$\alpha$ reported in the previous theoretical works when compared to the 
experimental data.  A comprehensive overview of the experimental results 
indicates the value of 49.50 reported by Qiao and collaborators \cite{qiao-12}
based on the experimental data of Goebel and Hohm \cite{goebel-95}, we believe,
is robust and reliable. This observation is based on 
three important considerations. First, the Wolfsohn's three term expression
\cite{wolfsohn-33} used in Ref. \cite{qiao-12}, to calculate $\alpha$ from the 
frequency dependent
dipole polarizability $\alpha(\omega)$, is an improvement over the three
term Cauchy expansion used in Ref. \cite{goebel-95}. Second, the value
50.0(2.8) reported in Ref. \cite{goebel-95}, based on the refractive index 
data from the work of Cuthbertson and Metcalfe \cite{cuthbertson-1908}, is 
consistent with the results in Ref. \cite{goebel-95,qiao-12}. Finally, in the
recent work of Hohm and Thakker \cite{hohm-12}, using a fitting function
with second ionization energy and Waber-Cromer radius \cite{waber-65}
as parameters, they
arrive at the value of $\alpha$ for Cd as 50.72. This is very closed to the
experimental values and must be given weightage as the values of
$\alpha$ reported in Ref. \cite{hohm-12}, except for Hf, Pd and Hg, are in 
good agreement with the reliable theoretical and experimental results. 
So, there is consistency in the experimental, and semi-empirical results
reported in the literature. This indicates the genesis of the lower 
theoretical results in the previous works must lie within the theoretical 
means and methods employed. 

  Returning to the  wide variation in the theoretical results, the possible 
reason for this could be, as evident from Table. \ref{pol_group2b} Cd has
the largest value of $\alpha$ among the group IIb elements. In addition,
$Z$ of Cd lies in the domain where relativistic effects begin to have an 
importance. So, in Cd, the relativistic and electron-correlation effects are 
inter-related strongly, as a result the properties which depend on electron 
correlation effects are sensitive to the choice of the basis set. One 
indication of this is the difference between the Hartree-Fock and CCSD(T) 
results of the $\alpha$ from the relativistic computations. From 
Ref. \cite{seth-97}, this is found to be 17.12 which is larger than the 
corresponding values of 12.18 and 10.36 for Zn and Hg, respectively. This 
demonstrates the importance of the relativistic and correlation effects.

\begin{table*}
        \caption{Orbital energies of Hg obtained from GRASP2K 
                 \cite{jonsson-13} and Gaussian type orbitals in atomic units. 
                 The quantities $\Delta{E}_{\rm Br}$ and $\Delta{E}_{\rm Ueh}$ 
                 are the orbital energy corrections arising from the Breit 
                 interaction and Uehling potential, respectively. In the 
                 table, [x] represents multiplication by ${10^x}$. All the 
                 values are in atomic units (hartree).
                 }
        \label{orb_hg}
        \begin{ruledtabular}
        \begin{tabular}{lddddd}
              {Orbital} 
              & \multicolumn{1}{c}{\text{GRASP2K}}     & 
                \multicolumn{1}{c}{\text{DC}}          &
                \multicolumn{2}{c}{\text{$\Delta{E}_{\rm Br}$}}         &
                \multicolumn{1}{c}{\text{$\Delta{E}_{\rm Ueh}$}}   \\
            \hline
              & & & \multicolumn{1}{c}{\text{Present}}&
                \multicolumn{1}{c}{\text{Ref. \cite{lindroth-89a}}} &  \\
            \hline
            $1s_{1/2}$ & -3074.226\,002 & -3074.235\,257 &  10.963\,407  
                       &  10.96         &  -1.557\,141  \\
            $2s_{1/2}$ & -550.251\,032  & -550.254\,927  &  1.229\,461   
                       &  1.230         & -2.206\,016[-1]  \\
            $2p_{1/2}$ & -526.854\,793  & -526.857\,122  &  2.067\,249   
                       &  2.067         & -1.415\,186[-2]  \\
            $2p_{3/2}$ & -455.156\,786  & -455.159\,068  &  1.304\,845   
                       &  1.305         &  7.499\,950[-3]  \\
            $3s_{1/2}$ & -133.113\,168  & -133.116\,535  &  2.275\,130[-1]   
                       &  2.276[-1]     & -5.013\,463[-2]  \\
            $3p_{1/2}$ & -122.639\,005  & -122.640\,349  &  3.933\,351[-1]   
                       &  3.933[-1]     & -3.454\,750[-3]  \\
            $3p_{3/2}$ & -106.545\,242  & -106.546\,285  &  2.346\,877[-1]   
                       &  2.347[-1]     &  2.184\,390[-3]  \\
            $3d_{3/2}$ & -89.436\,975   & -89.440\,259   &  1.708\,149[-1]   
                       &  1.708[-1]     &  2.426\,999[-3]  \\
            $3d_{5/2}$ & -86.020\,282   & -86.023\,564   &  1.098\,651[-1]   
                       &  1.098[-1]     &  2.298\,568[-3]  \\
            $4s_{1/2}$ & -30.648\,324   & -30.649\,589   &  4.665\,828[-2]   
                       &  4.667[-2]     & -1.258\,914[-2]  \\
            $4p_{1/2}$ & -26.124\,024   & -26.123\,690   &  8.337\,968[-2]   
                       &  8.339[-2]     & -7.139\,890[-4]  \\
            $4p_{3/2}$ & -22.188\,555   & -22.188\,057   &  4.359\,830[-2]   
                       &  4.360[-2]     &  7.141\,810[-4]  \\
            $4d_{3/2}$ & -14.796\,757   & -14.797\,894   &  2.297\,811[-2]   
                       &  2.297[-2]     &  7.153\,100[-4]  \\
            $4d_{5/2}$ & -14.052\,597   & -14.053\,659   &  9.563\,165[-3]   
                       &  9.554[-3]     &  6.841\,500[-4]   \\
            $4f_{5/2}$ & -4.472\,939    & -4.472\,953    & -5.808\,097[-3]   
                       & -5.816[-3]     &  5.019\,090[-4]  \\
            $4f_{7/2}$ & -4.311\,769    & -4.311\,745    & -1.148\,315[-2]   
                       & -1.150[-2]     &  4.923\,107[-4]  \\
            $5s_{1/2}$ & -5.103\,103    & -5.103\,080    &  7.030\,344[-3]  
                       &  7.033[-3]     & -2.389\,679[-3]  \\
            $5p_{1/2}$ & -3.537\,946    & -3.537\,438    &  1.212\,951[-2]   
                       &  1.213[-2]     &  3.017\,200[-6]  \\
            $5p_{3/2}$ & -2.842\,014    & -2.841\,487    &  4.829\,281[-3]   
                       &  4.828[-3]     &  2.641\,881[-4]  \\
            $5d_{3/2}$ & -0.650\,063    & -0.649\,907    &  2.431\,914[-4]   
                       &  2.394[-4]     &  2.060\,225[-4]   \\
            $5d_{5/2}$ & -0.574\,649    & -0.574\,475    & -1.088\,398[-3]   
                       & -1.093[-3]     &  1.954\,800[-4]  \\
            $6s_{1/2}$ & -0.328\,036    & -0.327\,943    &  4.584\,067[-4]  
                       &  4.575[-4]     & -2.026\,796[-4]  \\
        \end{tabular}
        \end{ruledtabular}
\end{table*}


\subsection{Hg}

 The results of Hg deserve detailed discussions as the current work is
precursor to a refined recalculation of the Hg atomic EDM \cite{latha-09}. 
Like in the previous cases, the orbital
energies of Hg and corrections are listed in Table. \ref{orb_hg}. From the
table it is evident that the values of $\Delta\epsilon_{\rm Br}$ from the
current work are in excellent agreement with the results reported
in Ref. \cite{lindroth-89a}. One noticeable change in the trend of 
$\Delta\epsilon_{\rm Br}$ is the 
negative values of $\Delta\epsilon_{{\rm Ue}(4f_{5/2})}$  and
$\Delta\epsilon_{{\rm Ue}(4f_{7/2})}$. In comparison, 
$\Delta\epsilon_{\rm Br}$ is negative for $3d_{5/2}$ and $4d_{5/2}$ in 
Zn and Cd, respectively. The results seem to indicate that the outermost 
sub-shell with $j\geqslant5/2$ have negative $\Delta\epsilon_{\rm Br}$, which 
could be on account of the larger weight factor $(2j+1)$ associated with higher
$j$ in the exchange two-electron integrals. The reason behind this remark is, 
only the exchange integrals contribute to the $\Delta\epsilon_{\rm Br}$ in 
closed-shell atoms and ions.

  The Uehling potential corrections to the orbitals energies exhibit one
marked change compared to Zn and Cd. In Hg, the values of 
$\Delta\epsilon_{{\rm Ue}(mp_{1/2})}$ with $ m = 2, 3, 4$ are negative. A
similar result was reported for the case of Ra$^{2+}$ in our previous work 
on doubly ionized alkaline-earth-metal atoms \cite{chattopadhyay-13b}. There 
is, however, one minor but important difference. In the case of Ra$^{2+}$ the
$\Delta\epsilon_{\rm Ue}$ is negative for all the $p_{1/2}$ orbitals. Whereas
in Hg, $5p_{1/2}$ orbital, the outermost $p_{1/2}$ orbital, has positive
$\Delta\epsilon_{\rm Ue}$. We attribute this to the larger relativistic
effects in Ra$^{2+}$ due to the stronger nuclear potential. Coming to the
basis set parameters, the values we have chosen generates orbitals with 
energies within $ 10^{-4}-10^{-3}$ hartree of the numerical orbital energies.  

  The PRCC computations with excitations from all the core sub-shells of Hg
generate cluster amplitudes in excess of $10^7 $  when the basis size
is $\sim 160$. The computation of $\alpha$, then, requires thousands of
hours of compute time, and detailed studies on the convergence properties is
unfeasible (with our existing facilities). To mitigate this computational
conundrum we restrict the cluster amplitudes to excitations from the 
$(4-6)s, (4-5)p, (4-5)d$, and $4f$ core sub-shells. From the results listed
in Table. \ref{conv_pat}, the $\alpha$ of Hg converges to 33.499 with a 
basis size of 134 orbitals. 

 Among the previous theoretical results, three are based on coupled-cluster
theory, and we discuss these in some detail. Consider first the CCSD(T) 
results of Kello and Sadlej \cite{kello-95}, it is obtained with
a polarized basis set, and correlating the $5d^{10}6s^{2}$ electrons. 
So, it is effectively 12 electron coupled-cluster calculations with 
relativistic corrections through the mass-velocity operator. Their result is
lower than ours, and below the experimental data as well. They also mention 
that $\alpha$ decreases to 31.24 when the computations are done with larger
number of correlated electrons, namely,  $5s^25p^65d^{10}6s^{2}$. So, the
primary reason for the difference may be the form of the 
relativistic effects. The second result is based on the CCSD(T) work of Seth 
and collaborators \cite{seth-97} using a basis set generated with an 
optimized quasirelativistic pseudopotential \cite{schwerdtfeger-86}. 
Their result is close to the experimental value, but on the higher side. 
The estimate of the contributions from the triple excitation 
is 0.84, which is smaller than the value 1.43 listed in the work of 
Kello and Sadlej \cite{kello-95}. This indicates that the contribution from
the triple excitation depends on the nature of basis set and form of the 
effective interaction to account for relativistic corrections. This is 
perhaps not surprising as the electron correlation effects subsumed through
the cluster operators depend on the nature of the basis functions. The third
or the last previous work \cite{pershina-08} on $\alpha$ of Hg with CCSD(T) 
is the closest, in terms of theoretical approach, to our present work. The
computations are based on the Dirac-Coulomb Hamiltonian, and their result
is within the experimental uncertainty. In summary, there is a variation
in the trend of the previous CCSD(T) results. The first \cite{kello-95}
and second \cite{seth-97} reports values which are below and above all the 
experimental data, respectively. The result of the third work 
\cite{pershina-08} is consistent with the experimental results. It must also
be mentioned that all of these three previous works are based on finite
field method.

  In the present work, as mentioned earlier, we use the Dirac-Coulomb-Breit 
atomic Hamiltonian. So, the Breit interaction is an additional relativistic 
effect considered in the present work compared to the previous coupled-cluster 
works. We must, however, add that there are other relativistic effects like 
frequency dependent transverse photon interaction not included in the present 
work. Our result of 33.50 is close, but below the experimental uncertainty of 
the most recent work \cite{goebel-96a}. With the inclusion of perturbative
$\mathbf{T}_3^{(1)}$ we get 33.59, this improves the agreement with 
experimental data.  Among the previous works, the results 
based on QCISD \cite{schwerdtfeger-94} and CI-MBPT \cite{hachisu-08} are in 
very good agreement with our result. In the latter case an important point is,
the basis set is generated with $V^{N-1}$ potential. Whereas all the other
previous works and ours are with basis generated using $V^N$ potential. 
Considering that the results from the recent works 
\cite{seth-97,pershina-08,hachisu-08}, and the present work are with different
methods, the relative variance of the results ($\approx 0.6$\%) is low. This 
demonstrates the methods do consolidate important relativistic and many-body
effects correctly. From this we can infer that the basis set, and 
PRCC(T) theory used in the present work is well suited
for precision computation of properties like atomic electric dipole moment.

 The term wise contribution, as listed in Table. \ref{cont_term},
Hg exhibits a noticeable change in the trend. The NLO contribution arises from
$\mathbf{T}_1^{(1)\dagger}\mathbf{D}T_2^{(0)} + \text{H.c.}$, where as
it is 
$\mathbf{T}_1^{(1)\dagger}\mathbf{D}T_1^{(0)} + \text{H.c.}$ in Zn and Cd. We
attribute this to the electron-correlation effects associated with the
electrons in $5d$ shell, which enhances the cluster amplitude of 
$T_2^{(0)}$. This is also reflected in the pattern of the core sub-shell
contribution to the LO term, 
where there is a marked change in the trend compared to Zn and Cd. The 
contribution from the valence shell, $6s_{1/2}$, is $\approx 40$\% smaller
than the valence sub-shell contribution in Cd. However, the contribution 
from the next core sub-shell $5d_{5/2}$ is more than double of 
$4d_{5/2}$ in Cd. This is on account of the relativistic contraction of 
the $6s_{1/2}$ radial wavefunction. 
%
%
%

 For Hg, two experimental results are available in the literature. First
is based on the data of dielectric constant reported in Ref. 
\cite{wusthoff-36}, and the other is based on the recent experimental 
measurement of Goebel and Hohm \cite{goebel-96a}. The two results are in 
very good agreement. There is another result \cite{tang-08} derived from the 
experimental data of Ref. \cite{goebel-96a} using the three term expression
of Wolfsohn \cite{wolfsohn-33}. The reanalysis is in view of the findings 
in Ref. \cite{salek-05} and \cite{gaston-06}, which report the need 
for eight or more terms, compared to three in Ref. \cite{goebel-96a}, in 
the Cauchy expansion of frequency dependent polarizability to obtain 
converged moments.


\subsection{Uncertainty estimates}

  We have identified different sources of uncertainties in the present work.
These arise from various approximations at different stages of the 
RCC and PRCC computations. The first two sources of uncertainties are 
associated with the truncation of the basis set, and consideration of cluster 
operators up to $T_3^{(0)}$ in the RCC theory. These are, however, negligible
as we consider a basis set which gives converged results of $\alpha$. 
The third source of uncertainty is the incomplete consideration of 
$\mathbf{T}^{(1)}_3$ as we include it perturbatively. To estimate an upper
bound on this uncertainty, consider the case of Hg, where the 
contribution from perturbative $\mathbf{T}^{(1)}_3$ is $\approx 0.3$\%, and is
the largest among the three atoms studied. Since the perturbative treatment
is considering the most dominant term, we can assume an uncertainty of
$\approx 0.3$\% as the upper bound arising from the remaining contributions
from  $\mathbf{T}^{(1)}_3$.  The fourth source of uncertainty is the 
truncation in the expression of $\alpha$ in Eq. (\ref{exp_alpha}), in which 
we retain terms up to second order in cluster operators. In one of our previous 
works \cite{mani-10}, we have shown the contribution from the third and higher
order terms in cluster amplitudes is negligible. So, the uncertainty from 
this can also be neglected. The last two sources of uncertainties are 
associated with the frequency dependent Breit interaction, and violation of 
no-virtual-pair approximation. In our previous work \cite{chattopadhyay-14a}, 
we had estimated the upper bound on the contribution from frequency dependent 
Breit interaction to be 0.13\% for Ra. For the present work too, as Ra has 
higher $Z$ than Hg, we consider this as the upper bound on the uncertainty 
arising from frequency dependent Breit interaction. As the systems under study 
are neutral atoms the contribution from the latter, violation of 
no-virtual-pair approximation, is negligible. Combining these, we estimate
the uncertainty in the results of Zn and Cd to below 0.5\%. For Hg, an
additional source of uncertainty is the restriction of excitations from the 
core sub-shells $(4-6)s$, $(4-5)p$, $(4-5)d$ and $4f$ in the converged 
basis set. Based on the computations with smaller basis set, but with 
excitations from all the core sub-shells, the upper bound on the uncertainty 
of the Hg results is $1.0$\%.


\section{Conclusion}

  We have computed the $\alpha$ of Zn, Cd and Hg, the elements of the 
groupIIB, using PRCC and our results are in very good agreement with the
experimental data. Among the three elements, our result of Cd is of
significance as ours is the only theoretical result consistent with the 
experimental data. Based on the analysis of available experimental data, we 
conclude that $\alpha$ of Cd reported by Qiao and 
collaborators \cite{qiao-12} is reliable. We attribute the lower values 
reported in the previous theoretical works to the choice of basis set, and 
the interplay of relativistic corrections with electron correlation effects. 
This is in contrast to the case of Zn and Hg, where the electron 
correlation, and relativistic corrections are predominant effects, 
respectively.

 In the PRCC sector, we have considered the triple excitation cluster operator
through the dominant contribution from the perturbative $\mathbf{T}_3^{(1)}$, 
and included it in the computation of $\alpha$. This brings the level of
electron correlation effects, in terms of excited state, in PRCC theory
on par with the RCCSDT theory we have developed and used.
The present work is based on use of Dirac-Coulomb-Breit atomic Hamiltonian.
In addition, we also consider the corrections from the Uehling potential, 
the leading order term in the vacuum polarization effects. So, we incorporate 
relativistic effects, albeit incomplete, better than the previous theoretical 
works. The relativistic effects left out in the present work include
self-energy corrections, frequency dependent transverse photon interaction 
and Wichmann-Kroll potential. We shall examine these in detail in future
works, and may be essential to reduce the uncertainties to below 0.5\% in the 
properties calculations of high $Z$ elements like Hg.

 An important highlight associated with an integral part of the 
Hamiltonian we use, Breit interaction, is the orbital energy correction 
associated with it. Our results are in excellent agreement with the previous 
results we could find in the literature, that is for Hg. This, we consider,
as a reliable validation of our implementation of Breit interactions. 
In future works, we shall report the application of PRCC theory to 
one- and two-valence systems. For which we have reported the results with 
unperturbed RCCSD theory \cite{mani-10,mani-11}.


\begin{acknowledgements}
We thank Arko Roy and Kuldeep Suthar for useful discussions. The
results presented in the paper are based on the computations using the 3TFLOP
HPC Cluster at Physical Research Laboratory, Ahmedabad.
\end{acknowledgements}


\appendix


\section{}
\label{app_a}

  The angular factors of the terms in the linearized RCCSDT equation of
$T_1^{(0)}$ given in Eq. (\ref{lccsdt_s}). In the expressions, $j_i$s are the
total angular momenta of the orbitals, and the quantities $[j]$ represent 
$2j+1$.
\begin{eqnarray}
   \label{sing_red}
    A_1 & = & \delta_{j_q,j_b}(-1)^{j_q - j_b}                    
                  \nonumber \\
    A_2 & = & (-1)^{j_p - j_b + k_1}\frac{\delta_{j_b,j_q}        
               \delta_{j_a,j_p}} {\sqrt{[j_b][j_p]}}
                  \nonumber \\
    A_3 & = & \frac{1}{2}                                         
              (-1)^{j_b+j_q+j_c+j_p}\frac{\delta_{j_a,j_p}} 
              {[k_1]\sqrt{[j_p]}}
                  \nonumber \\
    A_4 & = & \frac{1}{2}                                         
              (-1)^{j_b+j_q+j_c-j_p}\frac{\delta_{j_a,j_p}}
              {\sqrt{[j_p]}}
              \left \{ \begin{array}{ccc} 
                         k_1 & j_b & j_q \\
                         k_2 & j_c & j_p
                       \end{array} \right \}
                  \nonumber \\
    A_5 & = & (-1)^{j_b + j_q+j_r+j_p} \frac{\delta_{j_a, j_p}}   
              {[k_1]\sqrt{[j_p]}}
                 \nonumber \\
    A_6 & = & (-1)^{j_r+j_p+j_q-j_b}\frac{\delta_{j_a,j_p}}       
              {\sqrt{[j_p]}}
              \left \{ \begin{array}{ccc}
                         k_1 & j_r & j_b \\
                         k_2 & j_q & j_p
                       \end{array} \right \} 
                  \nonumber \\
    A_7 & = & \frac{1}{2[k_1] \sqrt{[l_2]}}                       
              (-1)^{j_b + j_q + j_c + j_r}
                  \nonumber
\end{eqnarray}
\begin{eqnarray}
    A_8 & = & \frac{1}{2\sqrt{[l_2]}}                             
              (-1)^{-j_b+j_r+j_c+j_q}
              \left \{ \begin{array}{ccc}
                                       k_1 & j_b & j_r \\
                                       l_2 & j_c & j_q
                                     \end{array} \right \}  \nonumber
\end{eqnarray}


\section{}
\label{app_b}

  The angular factors of the terms in the linearized RCCSDT equation of
$T_2^{(0)}$ given in Eq. (\ref{lccsdt_d}).
\begin{eqnarray}
   B_1 & = & \frac{\delta_{j_b,j_r}}{\sqrt{[j_b]}}                
                  \nonumber \\
   B_2 & = & \frac{\delta_{j_c,j_q}}{\sqrt{[j_c]}}                
                  \nonumber \\
   B_3 & = & (-1)^{j_a+j_p+j_b+j+q}                               
             \frac{1} {\sqrt{[k]}}
             \left \{ \begin{array}{ccc}
                        k_1 & j_a & j_c \\
                        j_p & k_2 & k
                      \end{array} \right \}
             \left \{ \begin{array}{ccc}
                        k_1 & j_d & j_b \\
                        j_q & k   & k_2
                      \end{array} \right \}
                  \nonumber \\
   B_4 & = & (-1)^{j_a+j_p+j_b+j_q}\frac{1}{\sqrt{[k]}}           
             \left \{ \begin{array}{ccc}
                        k_1 & j_p & j_r \\
                        j_a & k_2 & k  
                      \end{array} \right \}
             \left \{ \begin{array}{ccc}
                        k_1 & j_s & j_q \\
                        j_b & k   & k_2
                      \end{array} \right \}
                  \nonumber \\
    B_5 & = & \sum_{k_1} (-1)^{j_c+k+j_r}                        
              \left \{ \begin{array}{ccc}
                          j_a & j_c & k_1 \\
                          j_r & j_p & k
                        \end{array}\right \}
                   \nonumber \\
    B_6 & = & (-1)^{j_a+j_p+j_b+j_q+l_1+k_1+k_2}[l_1]            
              \left \{ \begin{array}{ccc}
                         k_1 & j_b & j_c \\
                         j_q & k_2 & l_1
                       \end{array} \right \}
                  \nonumber \\
        &&    \times
              \left \{ \begin{array}{ccc}
                         k_1 & j_r & j_p \\
                         j_a & l_1 & k_2
                       \end{array} \right \}
                  \nonumber \\
    B_7 & = & \frac{1}{2[k]}                                     
              ((-1)^{j_r+k-j_c}
                  \nonumber \\
    B_8 & = & \frac{1}{2}                                        
              (-1)^{j_c+k_1+j_r}
              \left\{\begin{array}{ccc}  
                       j_c & j_q & k_1 \\
                       j_b & j_r & k
                     \end{array} \right \}
                  \nonumber \\
    B_9 & = & (-1)^{j_r-j_c+j_b+j_q}                             
              \left \{ \begin{array}{ccc}
                         j_b & j_s & l_2 \\
                         l_1 & k   & j_q 
                       \end{array} \right \} 
                  \nonumber \\
 B_{10} & = & (-1)^{j_c+j_s+j_b+j_q+k+l_1}                     
              \left \{ \begin{array}{ccc}
                         k_1 & j_c & j_s \\
                         l_1 & j_q & j_r
                       \end{array} \right \} 
              \left \{ \begin{array}{ccc}
                         j_b & j_s & l_2 \\
                         l_1 & k   & j_q
                       \end{array} \right \}
                  \nonumber \\
 B_{11} & = & \frac{1}{2[k]}                                    
              (-1)^{j_r-j_c+j_b+j_q+l_2}
              \left \{ \begin{array}{ccc}
                         j_q & j_d & l_2 \\
                         l_1 & k   & j_b
                       \end{array} \right \} 
                  \nonumber \\
 B_{12} & = & \frac{1}{2}                                       
              (-1)^{j_c+j_r+j_b+j_q+l_2}
              \left \{ \begin{array}{ccc}
                         k_1 & j_r & j_d \\
                         l_1 & j_b & j_c 
                       \end{array} \right \}
              \left \{ \begin{array}{ccc}
                         j_d & l_2 & j_q \\
                         k   & j_b & l_1 
                       \end{array} \right \} \nonumber 
\end{eqnarray}


\section{}
\label{app_c}
  The angular factors of the terms in the linearized RCCSDT equation of
$T_3^{(0)}$ given in Eq. (\ref{lccsdt_t}). It is to be that the expression
of $C_4$ includes a $9j$-symbol.

\begin{eqnarray}
    C_1 & = & (-1)^{j_b+j_q+l_2}                                
              \left \{ \begin{array}{ccc}
                         l_3 & j_q & j_s \\
                         j_b & l_1 & l_2
                       \end{array} \right \}
                  \nonumber \\
    C_2 & = & (-1)^{j_b+j_q+l_3+l_1}                            
                \left \{ \begin{array}{ccc}
                           l_3 & j_b & j_d \\
                           j_q & l_1 & l_2
                         \end{array}\right \}
                  \nonumber \\
    C_3 & = & (-1)^{j_s-j_d+l_1}\frac{1}{[l_1]}                 
                  \nonumber
\end{eqnarray}
\begin{eqnarray}
    C_4 & = & (-1)^{j_b+2j_q+j_s+l_1}[l]                        
                \left \{ \begin{array}{ccc}
                           m_1 & l_3 & m_2 \\
                           j_b & l_2 & j_q \\
                           j_s & l_1 & j_d 
                         \end{array}\right \}
                  \nonumber \\
    C_5 & = & (-1)^{j_a+j_p+j_b+j_q+l_3}[l_1][l_2]              
              \left \{ \begin{array}{ccc}
                         m_1 & j_s & j_a \\
                         j_p & l_1 & k  
                       \end{array}\right \}
                  \nonumber \\
        &&    \times
              \left \{ \begin{array}{ccc}
                         j_q & j_d & m_2 \\
                         k   & l_2 & j_b
                       \end{array}\right \}
              \left \{ \begin{array}{ccc}
                         m_2 & m_1 & l_3 \\
                         l_1 & l_2 & k  
                       \end{array}\right \}
                  \nonumber \\
    C_6 & = & (-1)^{j_d+j_s+l_1}                             
              \left \{ \begin{array}{ccc}
                         j_a & j_d & k   \\
                         j_s & j_p & l_1
                       \end{array} \right \}
                  \nonumber \\
    C_7 & = & (-1)^{j_a+j_p+j_b+j_q+k+m_2+l_2+l_3} [l_1][l_2]
              \left \{ \begin{array}{ccc}
                         m_1 & j_s & j_a \\
                         j_p & l_1 & k  
                       \end{array}\right \}
                  \nonumber \\
        &&    \times
              \left \{ \begin{array}{ccc}
                         j_b & j_t & m_2 \\
                         k   & l_2 & j_q
                       \end{array}\right \}
              \left \{ \begin{array}{ccc}
                         m_2 & m_1 & l_3 \\
                         l_1 & l_2 & k  
                       \end{array}\right \}
                  \nonumber \\
   C_8 & = & (-1)^{j_a+j_p+j_b+j_q+k+m_1+l_1+l_3}[l_1][l_2]  
             \left \{ \begin{array}{ccc}
                        m_1 & j_d & j_p \\
                        j_a & l_1 & k  
                      \end{array}\right \}
                  \nonumber \\
       &&    \times
             \left \{ \begin{array}{ccc}
                        j_q & j_e & m_2 \\
                        k   & l_2 & j_b
                      \end{array}\right \}
             \left \{ \begin{array}{ccc}
                        m_2 & m_1 & l_3 \\
                        l_1 & l_2 & k  
                      \end{array}\right \}  \nonumber
\end{eqnarray}


\section{}
\label{app_d}
  The angular factors of the terms in the linearized PRCC equation of
$\mathbf{T}_1^{(1)}$ given in Eq. (\ref{lprcc_s}).


\begin{eqnarray}
   {\cal A}_1 & = & \frac{\delta{(j_a, j_q})}{\sqrt{[j_a]}}      
                                \nonumber  \\
   {\cal A}_2 & = & \frac{\delta{(j_b, j_p})}{\sqrt{[j_b]}}      
                                \nonumber \\
   {\cal A}_3 & = & \frac{1}{\sqrt{3}} (-1)^{j_q - j_b + 1}      
                                \nonumber \\
   {\cal A}_4 & = & (-1)^{j_b + j_q + 1}                         
                    \left \{ \begin{array}{ccc}
                               j_b  & j_q & 1  \\
                               j_a  & j_p & k
                             \end{array}\right\}  
                                \nonumber \\
   {\cal A}_5 & = & \frac{1}{\sqrt{3}}                           
                    (-1)^{j_q - j_b + k_1}
                                \nonumber \\
   {\cal A}_6 & = & (-1)^{j_b + j_q + 1}                         
                    \left \{ \begin{array}{ccc}
                                j_a  & j_b & k \\
                                j_q  & j_p & 1
                             \end{array}\right\}  
                                \nonumber \\
   {\cal A}_7 & = & \frac{1}{\sqrt{[m_2]}}                       
                    (-1)^{j_q - j_b + j_a + j_p + m_1} 
                    \left \{ \begin{array}{ccc}
                                j_r  & j_a & m_1 \\
                                1    & m_2 & j_p
                             \end{array}\right\}
                                \nonumber \\
   {\cal A}_8 & = & (-1)^{j_a + j_p + j_b + j_q + m_1}           
                    \left \{ \begin{array}{ccc}
                                j_b  & k   & j_r \\
                                j_p  & m_2 & j_q
                             \end{array}\right\} 
                    \left \{ \begin{array}{ccc}
                                j_r  & j_a & m_1 \\
                                1    & m_2 & j_p
                             \end{array}\right\}
                                \nonumber \\
   {\cal A}_9 & = & \frac{1} {\sqrt{[m_2]}}                      
                    (-1)^{j_a - j_b + j_p + j_q + 1 + m_2} 
                    \left \{ \begin{array}{ccc}
                               j_p  & m_1 & j_c \\
                               m_2  & j_a & 1
                             \end{array}\right\}  
                                \nonumber 
\end{eqnarray}
\begin{eqnarray}
  {\cal A}_{10}&= & (-1)^{j_q + j_c + j_a + j_p + 1 + m_2}       
                    \left \{ \begin{array}{ccc}
                                k   & j_a & j_c \\
                                m_2 & j_q & j_b
                             \end{array}\right\}    
                    \left \{ \begin{array}{ccc}
                                j_b  & j_p & m_1 \\
                                1    & m_2 & j_a
                             \end{array}\right\}, \nonumber
\end{eqnarray}


\section{}
\label{app_e}

  The angular factors of the terms in the linearized PRCC equation of
$\mathbf{T}_2^{(1)}$ given in Eq. (\ref{lprcc_s}). 

\begin{eqnarray}
  {\cal B}_1 & = & [l_1](-1)^{j_a + j_p + 1 + l_2}                
                   \left \{ \begin{array}{ccc}
                              1   & j_p & j_r \\
                              j_a & l_2 & l_1
                            \end{array}\right\}
                                \nonumber  \\
  {\cal B}_2 & = & [l_1] (-1)^{j_a + j_p + l_1}                   
                   \left \{ \begin{array}{ccc}
                              1   & j_a & j_c \\
                              j_p & l_2 & l_1
                            \end{array}\right\} 
                                \nonumber \\
  {\cal B}_3 & = & [l_1] (-1)^{j_a + j_p + l_1}                   
                   \left \{ \begin{array}{ccc}
                              l_2  & j_p & j_r \\
                              j_a  & 1   & l_1
                            \end{array}\right\} 
                                \nonumber \\
  {\cal B}_4 & = & [l_1](-1)^{j_a + j_p + 1 + l_2}                
                   \left \{ \begin{array}{ccc}
                              l_2  & j_a & j_c \\
                              j_p  & 1   & l_1
                            \end{array}\right\} 
                                \nonumber \\
  {\cal B}_5 & = & \frac{1}{[l_1]}                                
                   (-1)^{j_r - j_c + l_1}
                                \nonumber \\
  {\cal B}_6 & = & [l_2] (-1)^{j_c - j_q + 1 + m_2}               
                   \left \{ \begin{array}{ccc}
                              j_r & m_1  & j_b \\
                              j_c & m_2  & j_q \\
                              l_1 &  1   & l_2 
                            \end{array}\right\}
                                \nonumber \\
  {\cal B}_7 & = & [l_1][l_2] (-1)^{j_a + j_p + j_b + j_q + 1}    
                   \left \{ \begin{array}{ccc}
                              k   & j_p & j_r \\
                              j_a & m_1 & l_1
                            \end{array}\right\}             
                                \nonumber \\
             &&    \times
                   \left \{ \begin{array}{ccc}
                              l_2 & l_1 & 1  \\
                              m_1 & m_2 & k
                            \end{array}\right\}
                   \left \{ \begin{array}{ccc}
                              k   & j_c & j_b \\
                              j_q & l_2 & m_2
                            \end{array}\right\}
                                \nonumber \\
  {\cal B}_8 & = & (-1)^{j_c + j_r + l_1}                         
                   \left \{ \begin{array}{ccc}
                              j_a & j_c & k \\
                              j_r & j_p & l_1
                            \end{array}\right\} 
                                \nonumber \\
  {\cal B}_9 & = & [l_1][l_2]                                     
                   (-1)^{j_a + j_p + j_b + j_q + 1 + k + l_2 + m_2}
                   \left \{ \begin{array}{ccc}
                              k   & j_p & j_r \\
                              j_a & l_1 & m_1
                            \end{array}\right\}
                                \nonumber \\
             &&    \times
                   \left \{ \begin{array}{ccc}
                              m_1 & l_1 & k \\
                              l_2 & m_2 & 1
                            \end{array}\right\}
                   \left \{ \begin{array}{ccc}
                              k   & j_s & j_q \\
                              j_b & m_2 & l_2
                            \end{array}\right\}
                                \nonumber \\
  {\cal B}_{10}&=& [l_1][l_2]                                     
                   (-1)^{j_a + j_p + j_b + j_q + 1 + k + m_1 + l_1}
                   \left \{ \begin{array}{ccc}
                              k   & j_a & j_c \\
                              j_p & m_1 & l_1
                            \end{array}\right\}
                                \nonumber \\
             &&    \times
                   \left \{ \begin{array}{ccc}
                              l_2 & l_1 & 1  \\
                              m_1 & m_2 & k
                            \end{array}\right\}
                   \left \{ \begin{array}{ccc}
                              k   & j_d & j_b \\
                              j_q & l_2 & m_2
                            \end{array}\right\}       \nonumber
\end{eqnarray}


\bibliography{references}{}
\bibliographystyle{apsrev4-1}

\end{document}